\documentclass[aps,twocolumn,floats,superscriptaddress,nofootinbib,prl,reprint]{revtex4-1}

\usepackage{graphicx}
\usepackage{amsmath}

\begin{document}

\title{A framework for studying behavioral evolution by reconstructing ancestral repertoires}

\author{Dami\'{a}n G. Hern\'{a}ndez}
\email{D.G.H. and C.R. contributed equally to this work.}
\affiliation{Department of Physics, Emory University}
\affiliation{Department of Medical Physics, Centro At\'{o}mico Bariloche and Instituto Balseiro}
\author{Catalina Rivera}
\email{D.G.H. and C.R. contributed equally to this work.}
\affiliation{Department of Physics, Emory University}
\author{Jessica Cande}
\affiliation{Janelia Research Campus}
\author{Baohua Zhou}
\affiliation{Department of Physics, Emory University}
\affiliation{Department of Molecular, Cellular and Developmental Biology, Yale University}
\author{David L. Stern}
\affiliation{Janelia Research Campus}
\author{Gordon J. Berman}
\email{To whom correspondence should be addressed: gordon.berman@emory.edu.}
\affiliation{Department of Physics, Emory University}
\affiliation{Department of Biology, Emory University}

\date{\today}

\begin{abstract}
Although extensive behavioral changes often exist between closely related animal species, our understanding of the genetic basis underlying the evolution of behavior has remained limited. Here, we propose a new framework to study behavioral evolution by computational estimation of ancestral behavioral repertoires. We measured the behaviors of individuals from six species of fruit flies using unsupervised techniques and identified suites of stereotyped movements exhibited by each species. We then fit a Generalized Linear Mixed Model to estimate the suites of behaviors exhibited by ancestral species, as well as the intra- and inter-species behavioral covariances.  We found that much of intraspecific behavioral variation is explained by differences between individuals in the status of their behavioral hidden states, what might be called their ``mood." Lastly, we propose a method to identify groups of behaviors that appear to have evolved together, illustrating how sets of behaviors, rather than individual behaviors, likely evolved. Our approach provides a new framework for identifying co-evolving behaviors and may provide new opportunities to study the genetic basis of behavioral evolution.
\end{abstract}

\maketitle

Behavior is one of the most rapidly evolving phenotypes, with notable differences even between closely-related species \cite{cid,martins1996phylogenies}.  Variable behaviors and rapid behavioral evolution likely allows species to adapt rapidly to new or varying environments \cite{baier2019genetics,west2003developmental}.  Despite the importance of animal behavior, progress in revealing the genetic basis of behavioral evolution has been slow \cite{gleason2004quantitative,yamamoto2013genetic,ellison2011genetics,shaw2009genomic}.  In contrast, recent decades have seen significant progress in understanding the genetic causes of morphological evolution \cite{williams2009,shubin2009deep,levine2005gene,Stern:2013gv}.  

While there are many potential reasons for the discrepancy between studies of behavioral and morphological evolution, including the lack of a fossil record for behavior, a key difficulty is identifying which aspects of an animal's development and physiology are responsible for the observed changes in animals' actions. Changes in behavior along a phylogeny could emerge from alterations in the developmental patterning of neural circuitry (e.g., brain networks, descending commands, central pattern generators), hormonal regulation that affects the expression of behaviors, or the gross morphology of an animal's body or limbs \cite{Baker:2001gp}.  Each of these possibilities could result in behavioral effects at different, yet overlapping, scales -- from muscle twitches to stereotyped behaviors to longer-lived states like foraging or courtship or aging that may control the relative frequency of a given behavior -- making it difficult to identify the precise aspects of behavior that are changing.

To address these difficulties, the standard approach in the study of behavioral evolution has been to identify focal behaviors that exhibit robust differences between species, such as courtship behavior in fruit flies \cite{Jessica2012,Cande2014,YunDing2019} or burrow formation in deermice \cite{Hoekstra2013,Hu:2017:Seminar}.  With these robust differences in phenotype, it is possible to perform analyses that isolate regions of the genome that correlate with quantitative changes in the performance of the focal behavior. However, there tend to be multiple such regions identified, each containing many genes.  Given the large number of putative genes involved, combined with the possibility of epistatic interactions between loci, identification of the contributions of individual genes to behavioral evolution has moved slowly.

\begin{figure*}[ht]
\centerline{\includegraphics[width=.8\textwidth]{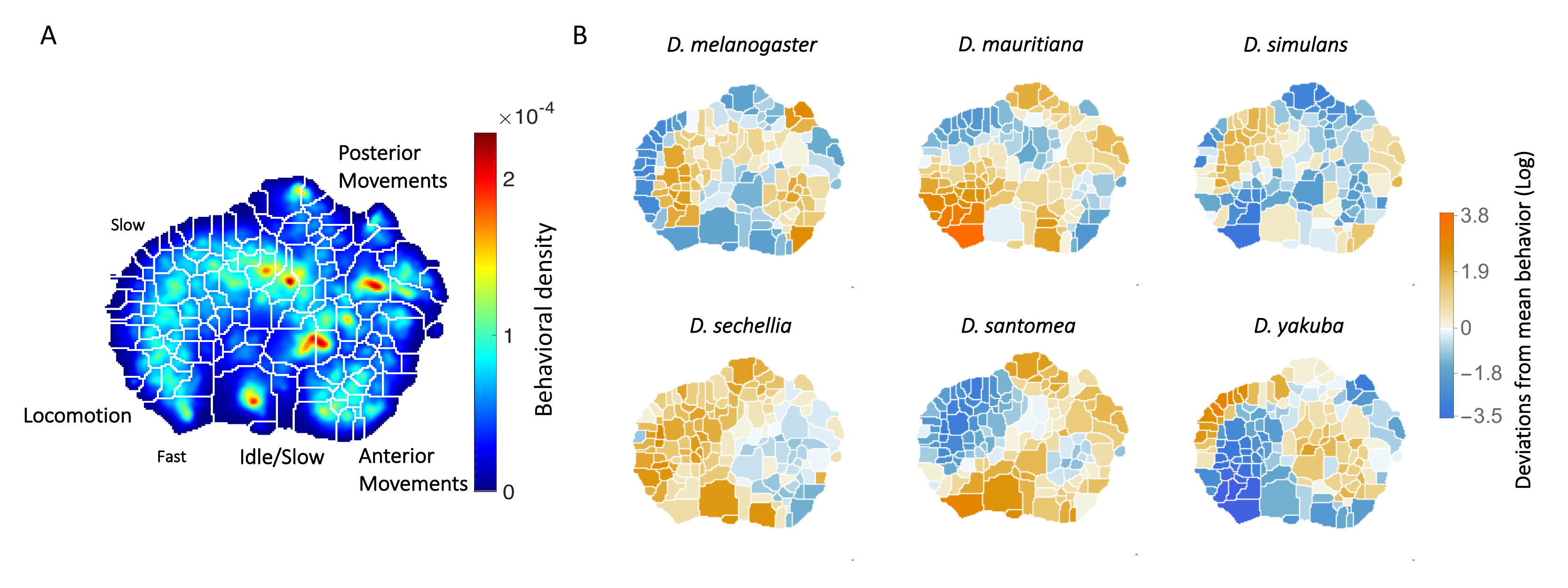}}
\caption{Behavioral repertoires of \emph{Drosophila}. A: The behavioral space probability density function, obtained using the unsupervised approach described in \cite{berman2014mapping} on the entire data set of $561$ individuals across all species. Coarse grained behaviors corresponding to the different types of movements exhibited in the map are shown as well. B: The relative performance of each of the 134 stereotyped behaviors for each of the six species.  Each region here represents a behavior, and the color scale indicates the logarithm of the fraction of time that species performs the specified behavior divided by the average across all species.}
\label{Fig1}
\end{figure*}

An alternative approach to focusing on single behaviors is to examine the full repertoire of movements that an animal performs.  By identifying sets of behaviors that evolve together, it may be possible to identify regulators of these suites of behaviors. This approach has been made possible by recent progress in unsupervised identification of animal behaviors across length and time scales \cite{GordonReview,Brown:2018ck}.  In this study, we introduce a quantitative framework for studying the evolutionary dynamics of large suites of behavior. We have focused initially on fruit flies, which provide a convenient model for this problem because they exhibit a wide range of complex behaviors and unsupervised approaches can be used to map all of the animal movements captured in video recordings \cite{berman2014mapping,cande2018optogenetic,Berman18102016}.

We recorded movies of isolated male flies from six species in a nearly stimulus-free environment. Because we did not record flies experiencing social and other environmental cues, we did not observe many charismatic natural behaviors, such as courtship and aggression. Nevertheless, we found that the behaviors they performed, including walking and grooming, contain species-specific information.  We thus hypothesized that our quantitative representations of behaviors could be studied in an evolutionary context. To infer the evolutionary trajectories of behavioral evolution, we estimated ancestral behavioral repertoires with a Generalized Linear Mixed Model (GLMM) approach \cite{Hadfield2010}, which builds upon Felsenstein's approach to reconstructing ancestral states \cite{Felsenstein:1985,y4r}. Using these results, we develop a framework that allows us to model the behavioral traits that co-vary both within a species and along the phylogeny. We find that within-species variance is related primarily to long-lasting internal states of the animal, what might be called a fly's “mood,” and that inter-species variance can capture how disparate behaviors may evolve together. This latter finding points towards the presence of higher-order behavioral traits that may be amenable to further evolutionary and genetic analysis.

\section*{Experiments and behavioral quantification}

We captured video recordings of all behaviors performed by single flies isolated in a largely featureless environment for multiple individuals from six species of the \emph{Drosophila melanogaster} species subgroup: \emph{D.~mauritiana}, \emph{D.~melanogaster}, \emph{D.~santomea}, \emph{D.~sechellia}, \emph{D.~simulans}, and \emph{D.~yakuba} \cite{cande2018optogenetic}.  Although the animals could not jump or fly in these chambers and were not expected to exhibit social or feeding behaviors, the flies displayed a variety of complex behaviors, including locomotion and grooming. Each of these behaviors involves multiple body parts that move at varying time scales. The species studied here were chosen because their phylogenetic relationships are well understood \cite{Clark:2007gf,Obbard:2012bk,chyb2013atlas,4z} (summarized in the tree seen in Fig.~\ref{Fig3}), and genetic tools are available for most of these species \cite{Stern:2017:G3}.  Since a single strain represents a genomic “snapshot” of each species, we assayed individuals from multiple strains from each species to attempt to capture species-specific differences, and not variation specific to particular strains (see Materials and Methods). In total, we collected data from 561 flies, each measured for an hour at a sampling rate of 100 Hz.

\begin{figure}[tb]
\centering
\includegraphics[width=0.4\textwidth]{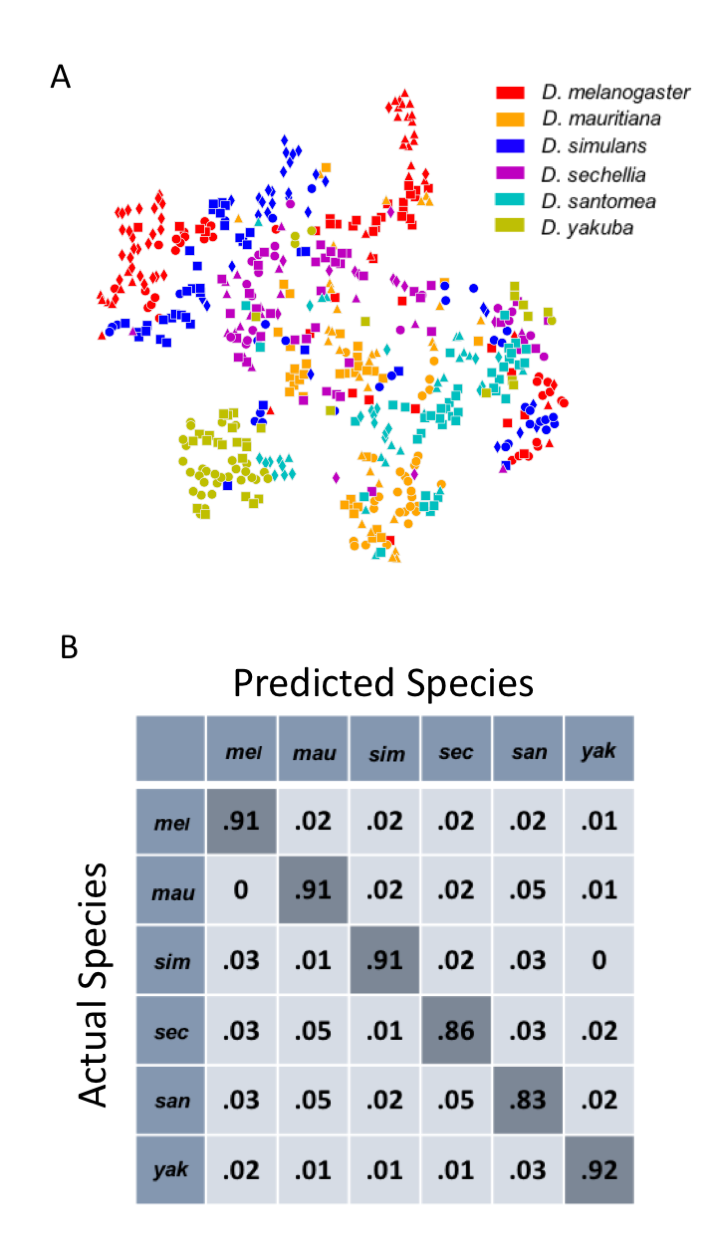}
\caption{Classification of fly species based on behavioral repertoires. A: A t-SNE embedding of the behavioral repertoires shows that behavioral repertoires contain some species-specific information. Each dot represents one individual fly, with different colors representing different species and different symbols with the same color representing different strains within the same species. The distance matrix (561 by 561) used to create the embedding is the Jensen-Shannon divergence between the behavioral densities of individual flies. B: Confusion matrix for the logistic regression with each row normalized. All the values are averaged from 100 different trials. The standard error is less than 0.01 for the diagonal elements and less than 0.005 for each of the off-diagonal elements.}
\label{Fig2}
\end{figure}

While previous studies have identified differences in specific behaviors, such as courtship behavior, between these species \cite{Jessica2012,YunDing2019,6cr,Auer:2016jc}, here we  assayed the full repertoire of behaviors the flies performed in the arena, with the aim of identifying combinations of behaviors that may be evolving together. To measure this repertoire, we used a previously-described behavior mapping method \cite{berman2014mapping, cande2018optogenetic} that starts from raw video images and attempts to find each animal’s stereotyped movements in an unsupervised manner. The output of this method is a two-dimensional probability density function (PDF) that contains many peaks and valleys (Fig.~\ref{Fig1}A), where each peak corresponds to a different stereotyped behavior (e.g., right wing grooming, proboscis extension, running, etc).

Briefly, to create the density plots, raw video images were rotationally and translationally aligned to create an egocentric frame for the fly. The transformed images were decomposed using Principal Components Analysis into a low-dimensional set of time series. For each of these postural mode time series, a Morlet wavelet transform was applied, obtaining a local spectrogram between 1 Hz and 50 Hz (the Nyquist frequency). After normalization, each point in time was mapped using t-SNE \cite{maaten2008visualizing} into a two dimensional plane. Finally, convolving these points with a two-dimensional gaussian and applying the watershed transform \cite{meyer1994topographic}, produced $134$ different regions, each of these containing a single local maximum of probability density that corresponds to a particular stereotypical behavior.  Thus, by integrating the density of the region for a particular fly, we can associate to each of them a $134$-dimensional real-valued vector that represents the probability of the fly performing a certain stereotyped behavior at a given time. We will refer to this quantity as the animal's \emph{behavioral vector} $\vec{P}$.

The behavioral map averaged across all six species is shown in Fig.~\ref{Fig1}A and displays a pattern of  movements similar to those we found in previous work, where locomotion, idle/slow, anterior/posterior movements, etc. are segregated into different regions  \cite{berman2014mapping,cande2018optogenetic}. Averaging across all individuals of each species, we found the mean behavioral vector for each species (Fig.~\ref{Fig1}B) and observed that each species performs certain behaviors with different probabilities. For example, \emph{D. mauritiana} individuals spend more time performing fast locomotion than all other species on average, and \emph{D. yakuba} individuals spend much of their time performing an almost species-unique type of slow locomotion, but little time running quickly.

These average probability maps provide some insight into potential species differences, but to identify species-specific behaviors, we also need to account for variation in the probability that individuals of each species perform each behavior. One way to address this problem is to ask whether an individual's species identity can be predicted solely from its multi-dimensional behavioral vector. To  explore this question, we first used t-SNE to project all 561 individuals into a 2 dimensional plane (Fig.~\ref{Fig2}A), using the Jensen-Shannon divergence as the distance metric between individual behavioral vectors. In this plot, different colors represent different species, and different symbols with the same color represent different strains within the same species.   Although there is not a clear segregation of all species in this plane, the distribution of species is far from random, with individuals from the same species tending to group near to each other.

To quantify this observation, we  applied a multinomial logistic regression classifier that performed a six-way classification based solely on the high-dimensional behavioral vectors. After training, the classifier correctly classified $89 \pm .2\%$ of vectors (using a randomly-selected test set of 30\% of the entire data set).  Moreover, the confusion matrix (Fig.~\ref{Fig2}B) revealed no systematic misclassifications bias amongst the species.  Note that we have used a relatively simple classifier compared to modern deep learning methods \cite{goodfellow2016deep}, so these results likely represent a lower bound on the distinguishability of the behavioral vectors. Thus, behavioral vectors appear to contain considerable species-specific information. We therefore proceeded to explore how these behavioral vectors may have evolved along the phylogeny.

\begin{figure*}[t]
\centerline{\includegraphics[width=.7\textwidth]{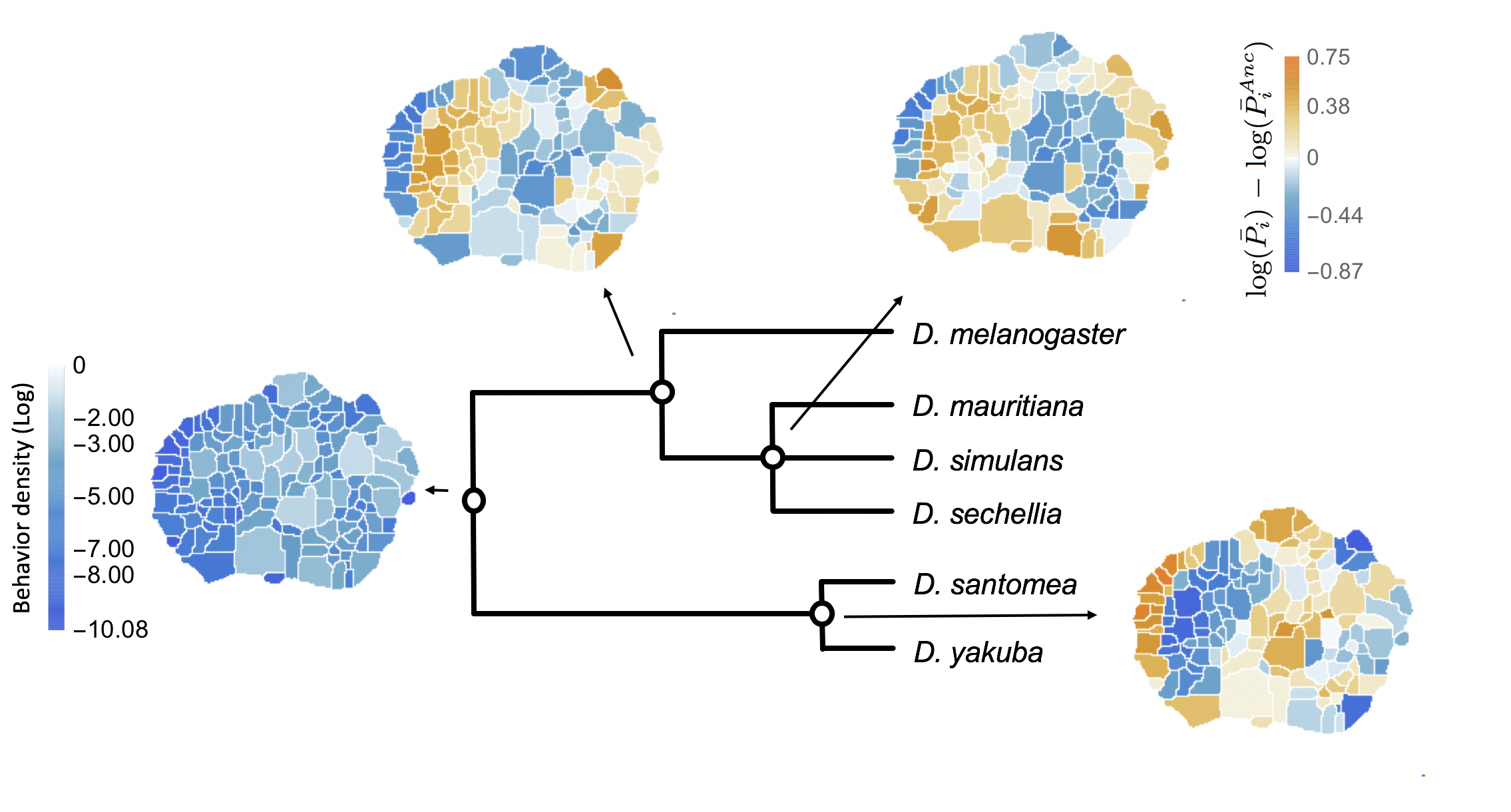}}
\caption{Reconstructed behavioral repertoires using the GLMM. Inferred probabilities of the behavioral traits for the ancestral states are plotted in logarithmic scale. Except for the ancestral root, other ancestral states are plotted with respect to the closest ancestor. Here, all but the root ancestor are plotted with respect to their closest ancestral state.  Therefore, for each behavioral trait, $i$,  we show: $\log(\bar{P}_{i})-\log (\bar{P}^{Anc}_{i})$, where $\bar{P}{i}$ and $\bar{P}^{Anc}_{i}$ correspond to the inferred mean behavioral trait for the given ancestor and its closest ancestor, respectively. }
\label{Fig3}
\end{figure*}

\section*{Reconstructing Ancestral Behavioral Repertoires}

Multiple methods have been proposed for reconstructing ancestral states solely from data collected from extant species \cite{Felsenstein:1985,yang2006computational}. These methods generally fall into two camps: parsimony reconstruction, which attempts to reconstruct evolutionary history with the fewest number of evolutionary changes \cite{md}, and diffusion-processes, which model evolution as a random walk on a multi-dimensional landscape \cite{hadfield2010general}. Given the high-dimensional behavioral vectors that we are attempting to model, a diffusion process is more likely to capture the inter-trait correlations that we would like to understand. Thus, we focus on a diffusion-based model here.

Given a phylogeny for a collection of species, we modeled how species-specific complexes of behaviors might have emerged. Specifically, we assumed that each behavior is a quantitative trait, that is, each behavioral difference results from the additive effects of many genetic loci, each of small effect. We do not, however, assume that all behaviors evolve independently of each other.  Thus, we are interested in predicting (1) how behaviors co-vary and (2) whether intra- and inter-species variation can be separated to identify independently evolving sets or linear combinations of behaviors.

We assumed that the flies' behaviors evolved via a diffusion process, where initially the process starts at the common ancestor behavioral representation and eventually each individual's trajectory performs a random walk with Gaussian noise along the known phylogenetic tree.  Note that this is a less stringent assumption than neutrality, as multiple traits under selection may evolve in a correlated manner.  More precisely, we fit a Multi-response Generalized Linear Mixed Model (GLMM) to the data, using the approach described in  \cite{Hadfield2010}:
\begin{equation}
\vec{l}= \vec{\mu} + \vec{\rho} + \vec{e}
\label{Model1}
\end{equation}
where $\vec{l}=(l_{1},..., l_{K=134})$ denotes the logarithm of the \emph{behavioral vector} $\vec{P}$ for each individual, $\vec{\mu}$ is the mean behavior of the common ancestor (treated as the fixed effects of this model), and $\vec{\rho}$ and $\vec{e}$ are the random effects corresponding to the phylogenetic and individual variability, respectively.  We assume that these random effects are generated from the multi-dimensional normal distributions $\mathcal{N}(\vec{0}, A\otimes V^{(a)})$ (phylogenetic) and $\mathcal{N}(\vec{0}, I\otimes V^{(e)})$ (individual). Here, the matrix $A$ represents the information contained in the phylogenetic tree, with $A_{ij}$ being proportional to the length of the path from the most recent common ancestor of species $i$ and $j$ to the main ancestor. This matrix is normalized so that the diagonal elements are all equal to $1$. $I$ is the identity matrix, and $V^{(a)}$ and $V^{(e)}$ are the covariance matrices that govern the process.  We fit $\mu$, $V^{(a)}$, and $V^{(e)}$ using Markov Chain Monte Carlo (MCMC) simulation (see Materials and Methods). We checked that the MCMC converged using the Gelman-Rubin diagnostic (see Materials and Methods, Fig.~\ref{RubinGelman}).  In addition to the inferred behavioral states corresponding to the common ancestor, $\bar{P}^{Anc}$, we also reconstructed the mean behavioral representations for the intermediate ancestors (Fig.~\ref{Fig3}). Further validation of our results corresponding to the current species behavior is shown in Fig. \ref{RubinGelman2}.

\begin{figure*}[t]
 \centering
\centering
\includegraphics[width=14.0cm]{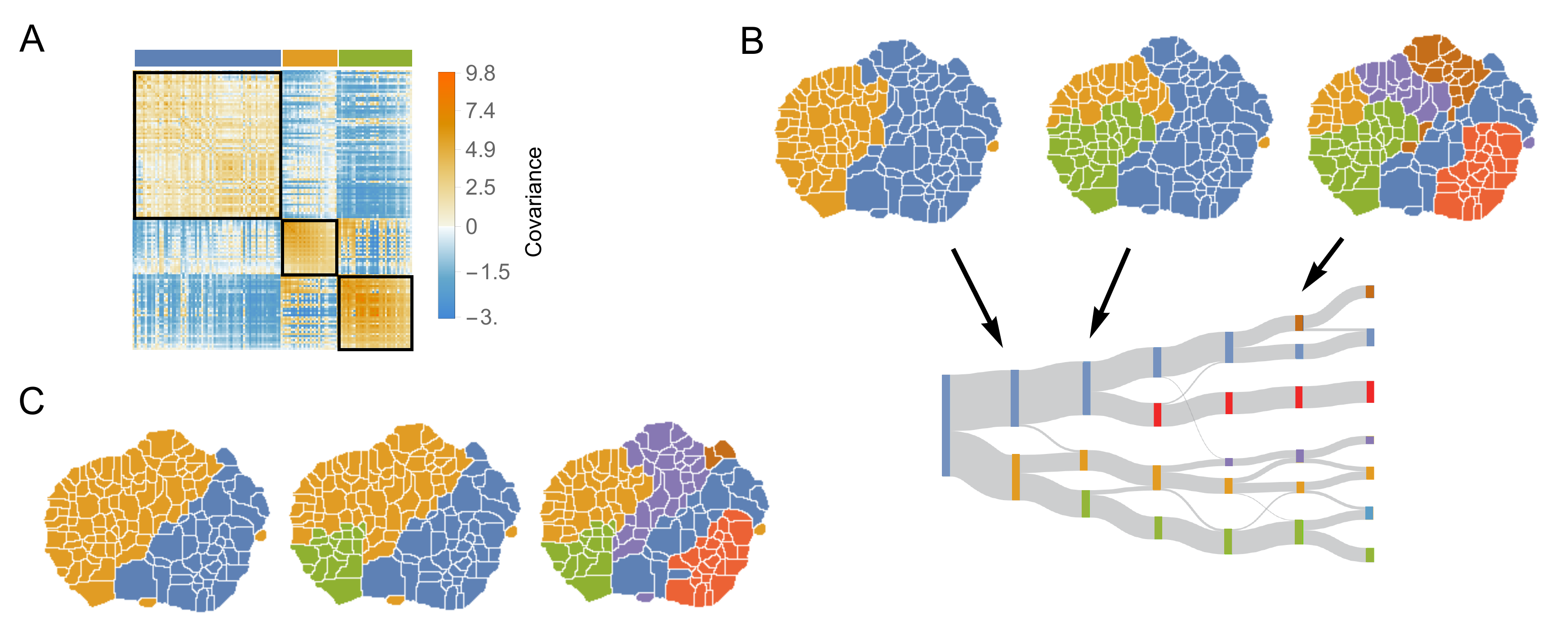}
\caption{The structure of variability between flies of the same species relates to long timescale transitions in behavior. A: The intra-species behavioral covariance matrix ($V^{(e)}$), with columns and rows ordered via an information-based clustering algorithm \cite{Slonim20122005}. The black squares represent behaviors that are grouped together in the three cluster solution.  B: Behavioral map representation of the clustering solutions.  The two, three, and six cluster solutions are shown on top (colors on the three cluster solution match those above the plot in A).  The clusters are all spatially contiguous and break down hierarchically (see Fig.~\ref{ClusterTypes} for more examples).  C: Clustering structure of the behavioral space obtained finding the optimally predictive groups of behaviors (see text for details).  Note how these clusterings are nearly the same as the clusterings in B, despite having been derived from an entirely independent measure.}
\label{fig:ind}
 \end{figure*}

\begin{figure*}[t]
 \centering
\includegraphics[width=14.0cm]{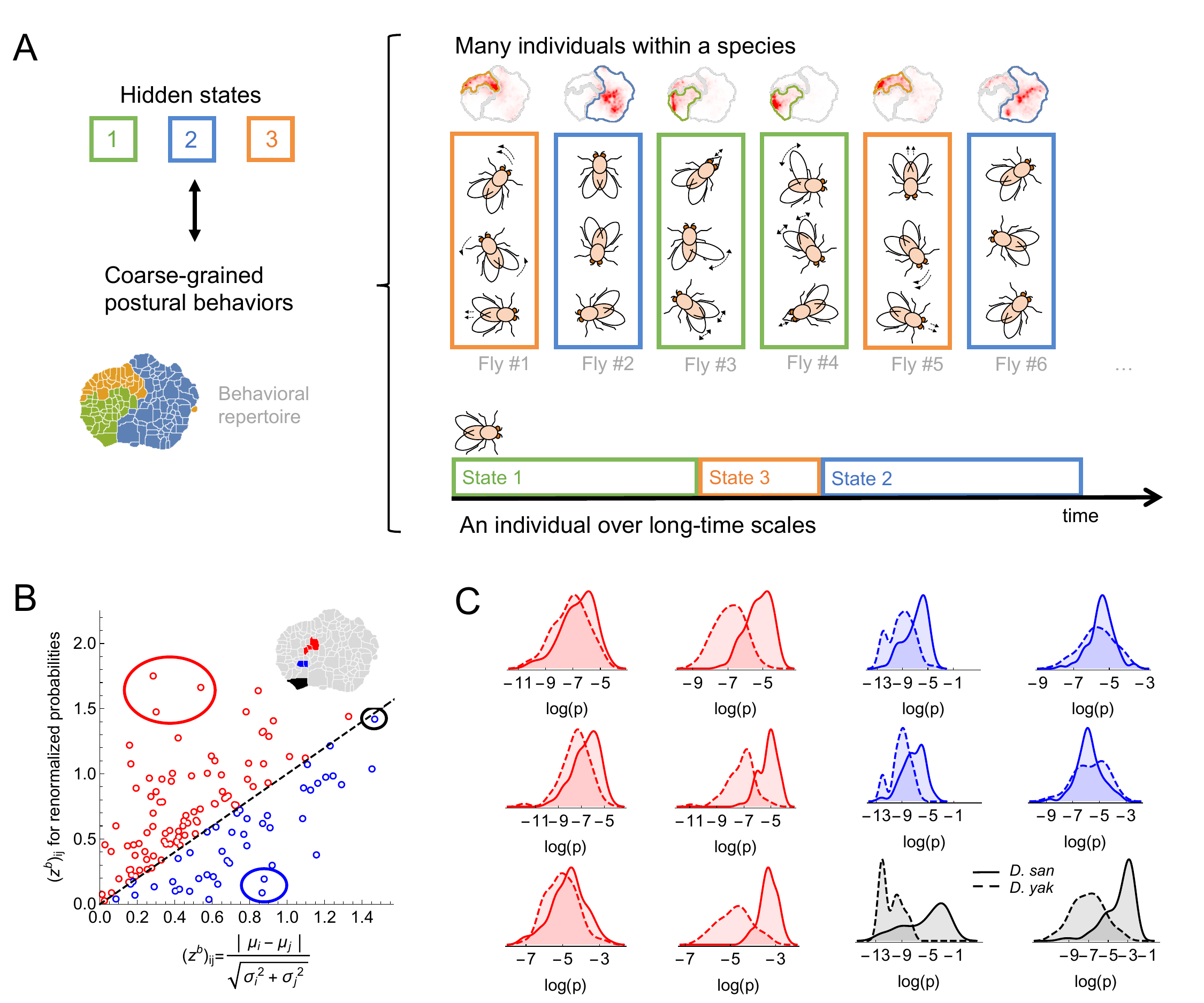}
\caption{Variability within a species, long timescale transitions, and hidden states modulating behavior.  A: A cartoon of the hypothesized relation between individual variability within a species and long timescale transitions through hidden states. B: Accounting for the long timescale dynamics - by adjusting for the amount of time spent in each coarse-grained region (here, the six cluster solution at the top right of Fig.~\ref{fig:ind}C) - affects the measured behavioral distributions between \emph{D. santomea} and \emph{D. yakuba}.  Shown is the comparison of the Mahalanobis distance ($(z^b)_{ij}$) between behavioral distributions before (x-axis) and after (y-axis) adjusting. C: Kernel density estimates of the distributions for the circled behaviors in B) on the left before (left) and after (right) adjustments. Solid lines represent \emph{D. santomea} and dashed lines represent \emph{D. yakuba}.}
\label{fig:ind2}
 \end{figure*}

\section*{Individual variability and long timescale correlations}

While it is not possible to directly test the accuracy of our ancestral state reconstructions, the inferred covariance matrices generate predictions about genetic correlations that are, in principle, testable. We therefore focus on our fitted covariance matrix, $V^{(e)}\in\Re^{134\times 134}$, which accounts for within-species random effects.  

We first note that $V^{(e)}$ exhibits a modular structure (Fig.~\ref{fig:ind}A).  After rearranging the behavior order via an information-based clustering procedure \cite{Slonim20122005}, we see that a block diagonal pattern emerges, with positive correlations lying within the blocks and negative correlations lying off the diagonal. This clustering approach minimizes the functional $\mathcal{F}=\langle d \rangle + \beta I(C;b)$, where $\langle d \rangle$ is the average within-cluster distance between behaviors (defined here as $d_{ij}=\frac{1}{2}[1-V^{(e)}_{ij}/\sqrt{V^{(e)}_{ii}V^{(e)}_{jj}}]$), $I(C;b)$ is the mutual information between cluster assignment and behavior number, and $\beta$ modulates the relative importance of the two terms (see Materials and Methods).  This modular structure emerges when applying other clustering methods as well (Fig.~\ref{ClusterTypes}).  Quantifying the matrix's modularity, we find that $\langle d \rangle \approx 0.30$ and $0.22$ for the 3 and 6-cluster solutions respectively. These values are significantly smaller than the average distances obtained using random cluster assignments ($\langle d \rangle= 0.46 \pm{0.03}$ and $0.45 \pm{0.04}$ for 3 and 6 clusters respectively, see Fig.~\ref{ShuffleInfo}).  Strikingly, these clusters are spatially contiguous in the behavioral map -- implying that similar behaviors explain much of the intra-species variance \cite{Berman18102016}. Moreover, new clusters emerge in a hierarchical fashion, where coarse-grained behaviors sub-divide into new clusters (Fig.~\ref{fig:ind}B), a feature that is not guaranteed by the information-based clustering algorithm.

This hierarchical structure of the behavioral space is reminiscent of the hierarchical temporal structure of behavior that was hypothesized originally by ethologists \cite{tinbergen51} and was observed to optimally explain the long timescale structure of \emph{Drosophila melanogaster} behavioral transitions \cite{Berman18102016}. To explore this connection further, we found coarse-grainings of the behavioral space that are optimally predictive of the future behaviors that the flies perform via the Deterministic Information Bottleneck (DIB) \cite{strouse2017deterministic}. Similar to the previously described information-based clustering method, this approach minimizes a functional, 
$\mathcal{J}_\tau= -I(b(t);Z(t+\tau)) + \gamma \mathcal{H}(Z)$, 
where $b(t)$ is a fly's behavior at time $t$, $Z(t+\tau)$ is the coarse-grained behavior visited at time $t+\tau$, $\tau=50$, $I(b(t);Z(t+\tau))$ is the mutual information between these quantities, $\gamma$ is a positive constant, and $\mathcal{H}(Z)$ is the entropy of the coarse-grained representation (see Material and Methods).  As $\gamma$ is increased, progressively coarser representations are found.

Applying this method to the data pooled across all six species (Figs.~\ref{fig:ind}C, \ref{ParetoFrontDIB}), we again found the same type of hierarchical division in the behavioral space that was observed for freely moving \emph{D. melanogaster} \cite{Berman18102016}.  Moreover, we found that the structure of the space using this approach closely mirrors the structure found via clustering $V^{(e)}$ (Fig.~\ref{fig:ind}C).  We quantify the similarity between both clustering partitions by calculating the Weighted Similarity Index (WSI), a modification of the Rand Index \cite{rand1971objective} (Materials and Methods). The WSI between the information-based clustering method and the predictive information bottleneck for three clusters is $WSI=0.73$ and $WSI=0.87$ for six clusters.  For random clusterings, we would expect to observe $0.51\pm 0.02$ and $0.70\pm 0.01$ for 3 and 6 clusters, respectively, indicating a non-random overlap between these two partitions. Fig.~\ref{ClusterTypes}, shows that this result is independent of the clustering method and the number of clusters. 

The overlap between these two coarse-grainings indicates that most individual variability in the behaviors we observe results from non-stationarity in behavioral measurements, rather than from individual-specific variation. That is, much of the intraspecific variation appears to reflect flies recorded when they were experiencing different hidden behavioral states (i.e. “moods”), rather than reflecting fixed (environmental or genetic) differences between flies. This variation may have arisen because, although we controlled many variables (e.g., fly age, circadian cycle, temperature, and humidity), it is not possible to control for all internal factors (e.g., hunger, arousal, etc.) that affect an animal’s behavioral patterns \cite{anderson2016circuit}.  The temporal coarse-graining of the behavioral space that we found via the DIB, gives insight into these non-stationarities, as they are optimally-predictive of the fly’s future behaviors.  Given the contiguous nature of these regions, this result means that flies tended to stay within specific regions of the behavioral space much longer than one would assume from a Markov model.

\begin{figure*}[t]
\centering
\includegraphics[width=14cm]{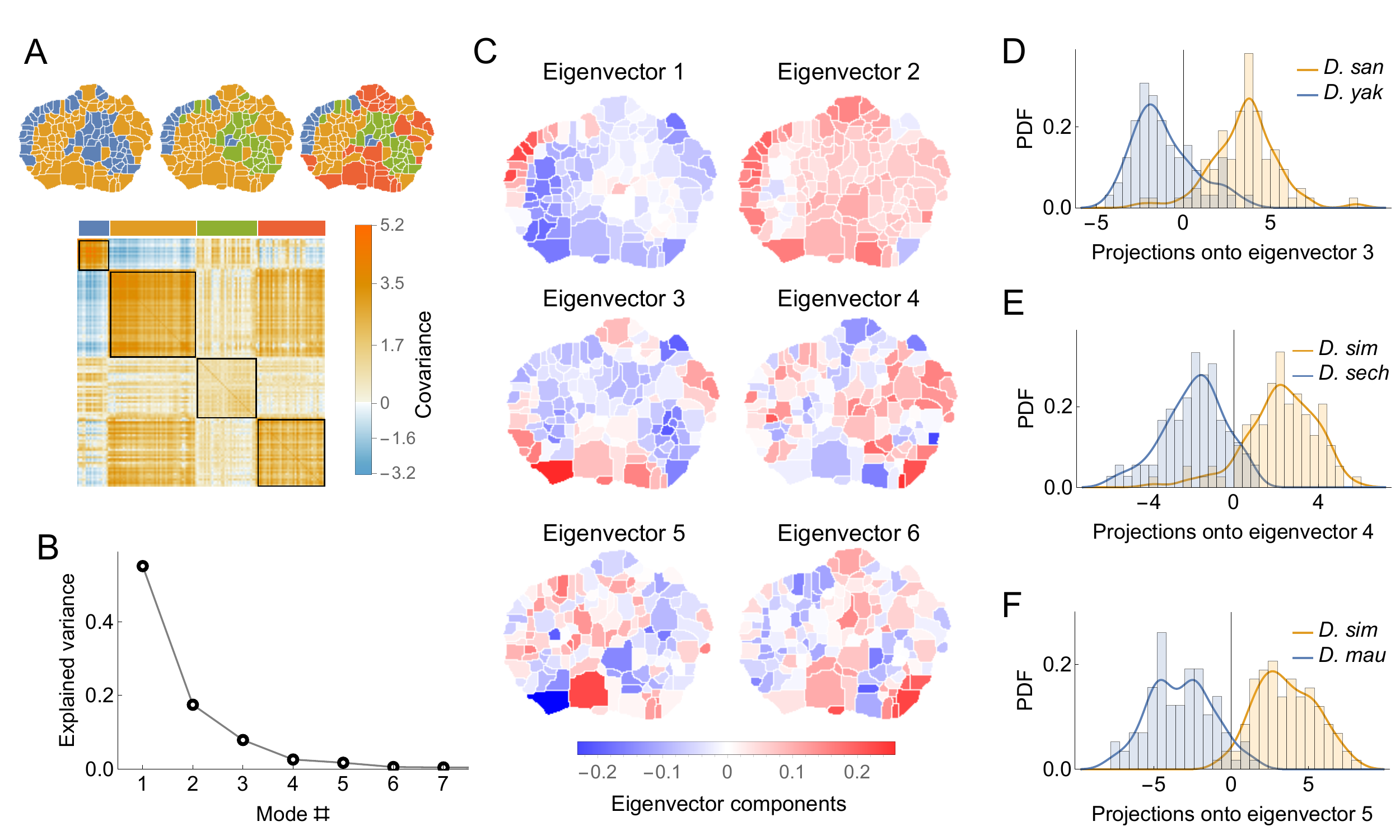}
\caption{Phylogenetic variability and behavioral meta-traits. A: (top) Clustering the phylogenetic covariance matrix (using the same information-based clustering method from Fig.~\ref{fig:ind}), we observe that the clusters are no longer spatially contiguous. (bottom) The phylogenetic covariance matrix reordered according to four clusters (colors corresponding to the four-cluster map above). B: Fraction of variance explained by the largest eigenvalues of the phylogenetic covariance matrix.  C: The eigenvectors corresponding to the largest six eigenvalues. D: Distributions of the projections of individual density vectors from \emph{D. santomea} and \emph{D. yakuba} onto eigenvector 3.
E: Same as in D but using projections of individuals from  \emph{D. sechellia} and \emph{D. simulans} onto eigenvector 4. F: Same as in D but using projections of individuals from  \emph{D. simulans} and \emph{D. mauritiana} onto eigenvector 5.}
\label{fig:side}
\end{figure*}

This observation implies that variation in behavior observed among individuals, especially in non-manipulated settings, is likely to often reflect a large component of hidden behavior states (Fig.~\ref{fig:ind2}A). Thus, it may be possible to improve upon behavioral measurements in many settings by controlling for the variability associated with these hidden states. For example, just because one fly performs less anterior grooming than another may reflect that the animal is in a different long timescale behavioral state, rather than that the animal has a genetically encoded preference for reduced grooming. 

A potential method for accounting for these artifacts is to normalize each individual's behavioral density such that the amount of time that the animal spends in each of the coarse-grained regions is equalized.  In other words, the amount of time spent anterior grooming, locomoting, etc. are set to be the same for all animals, thus accounting for the variability associated our inferred hidden states. Mathematically, if $P_i$ is the probability of observing behavior $i$, and $C_i$ is the clustering assignment of this behavior, we can define a normalized probability, $\hat{P}_i$, via 
\begin{equation}
    \hat{P}_i = \frac{\bar{P}^{(C_i)}}{P_i^{(C_i)}}P_i,
\end{equation}
where $P_i^{(C)} = \sum_{k\in C} P_k$ is the total density in cluster $C$ for an individual fly and $\bar{P}^{(C)}$ is the average across all animals.  

We found that applying this normalization to our data often results in substantial changes in the
inferred distributions of behavioral densities. For example, Fig.~\ref{fig:ind2}B displays how the difference in behavioral density between \emph{D.~santomea} and \emph{D.~yakuba} (as measured by the Mahalanobis distance between the distributions) alters as a result of normalization. For some behaviors, the signal increases (red points), and in some cases, it reverses (blue points). Thus, it is important to take these non-stationary effects into account when estimating how often single behaviors are performed in studies of behavioral evolution. To measure these non-stationary effects, many behaviors must be measured, not just a focal behavior. 

\section*{Identifying phylogenetically linked behaviors}

One of the advantages of our approach is that we separate variations in behavior corresponding to evolutionary patterns, the phylogenetic variability, from variations among individuals of the same species.  By studying the properties of the phylogenetic covariance matrix ($V^{(a)}$), we can identify behaviors that may be evolving together. 

We first characterized the coarse-grained structure within $V^{(a)}$ through the information-based clustering described in the previous section \cite{Slonim20122005}. As seen in Fig.~\ref{fig:side}A, these clusters are not spatially contiguous in the behavioral space. This pattern contrasts to the spatial contiguity we observed for the individual covariance matrix (Fig.~\ref{fig:ind}B). For example, the two-cluster solution (Fig.~\ref{fig:side}A, left) separates the behavioral space into side legs movements (middle) and certain locomotion gaits (far left) from the rest of behaviors. Similarly, non-localized structure is also observed when the matrix is clustered into a larger numbers of clusters as well.

One possible interpretation of these discontinuous clusters is that at the neural level, each of these groups of movements may reflect a motor response to shared upstream commands \cite{cande2018optogenetic}. For example, different types of locomotion might be controlled through the same descending neural circuitry, but due to evolutionary changes, the same commands could lead to different behavioral outputs, as has been observed in fly courtship patterns \cite{YunDing2019}. Thus, examination of phylogenetically course-grained regions such as these may provide a more biologically realistic view of suites of evolving behaviors than does focus on single behaviors. 

To quantify these patterns as traits, we decomposed $V^{(a)}$ via an eigendecomposition.  As seen in Fig.~\ref{fig:side}B, almost all of the variance within the matrix can be explained with only the first six eigenmodes.  These eigenvectors (Fig.~\ref{fig:side}C) share similar non-local structure to the clusterings described above.  By projecting individual behavioral vectors onto these eigenvectors, the resulting dot products represent a meta-trait that is a linear combination of phylogenetically linked behaviors. 

These evolving meta-traits may be suitable targets for further neurobiological or genetic studies. Three examples of these distributions are shown in in Fig.~\ref{fig:side}D for several pairs of closely related species. These three examples were not chosen at random, but instead because they showed significant differentiation between species. The aim of this analysis is not to show that all meta-traits would differ between all pairs of species, which strikes us as unlikely, but rather that it is possible to identify synthetic meta-traits that could be further interrogated with experimental methods.

\section*{Discussion}

We have developed a quantitative framework to study the evolution of behavioral repertoires, using fruit flies (\emph{Drosophila}) as a model system. We started with observations of 561 individuals from six extant species behaving in an unremarkable environment. This assay did not include social behaviors, such as courtship and aggression, nor many foraging behaviors. Thus, at first glance, it might seem like we had excluded most species-specific behaviors from the analysis. Nonetheless, we found that other complex behaviors, like walking, running, and grooming, exhibit species-specific features that can be used to reliably assign individuals to the correct species. Thus, the motor patterns of behaviors that are not normally investigated for their species-specific features are clearly evolving between even closely related species. It is not clear if these differences reflect natural selection or genetic drift on the details of these motor patterns. But, all of these behaviors would seem to be critical to individual survival, so it is possible that these behaviors have evolved, at least in part, in response to natural selection. It is clear, however, that the underlying neural circuitry controlling these behaviors must have evolved.

Inspired by these observations, we estimated patterns of behavioral evolution in the context of a well-understood phylogeny. We fit a Generalized Mixed Linear Model to our behavioral measurements and the given phylogeny to reconstruct ancestral behavioral repertoires and the intra- and inter-species covariance matrices. We found that the patterns of intra-species variability are similar to long timescale behavioral dynamics. This suggests that much of the intraspecific variability that emerged by sampling flies under well-controlled conditions reflects variability in the hidden behavioral states of individual flies. This variability is a clear confound for evolutionary and experimental studies of behavior and we therefore propose a method to control for these internal states and improve the accuracy of behavioral phenotyping.  We showed that controlling for these internal states can dramatically alter estimates of the “heritable” elements of behavior.
 
Given our estimates for how suites of behaviors evolved, we examined whether the inter-species covariance matrix could be used to identify behavioral meta-traits that might be subjected to further evolutionary and experimental analysis. We identified multiple suites of behaviors that differed between closely related species, providing a starting point for further analysis of how the mechanismns underlying these suites of behaviors have evolved.

The analysis framework introduced here represents the first attempt to analyze full behavioral repertoires to gain insight into evolution. In principle, this approach could be applied to any data set where a large number of behaviors have been sampled in many species. We envision several areas where future improvements may yield more detailed, comprehensive, and biologically meaningful results. First, we recorded behavior from only six species of flies. Adding additional species would place more constraints on the evolutionary dynamics, likely resulting in less variance in the ancestral state estimations and potentially adding more structure to the relatively low rank covariance matrices. Additionally, further work is required to determine the balance between sampling within and between strains and species that optimizes estimates of evolutionary dynamics.

Second, our framework assumes that all evolutionary changes in behavior resemble a diffusion process. Although this assumption is a reasonable initial hypothesis \cite{Felsenstein:1985}, it may be possible to test this assumption. For example, deeper sampling of additional species may allow identification of specific behaviors on particular lineages where neutrality can be rejected \cite{Tajima.1993.Genetics}.

In addition, all of our analyses involved measuring the fraction of behaviors performed during the recording time, ignoring the temporal structure and sequences of movements.  While we show here that much of this information can be related to the structure of the intra-species covariance matrix, the order in which behaviors occured may also provide important biological information.  It should be possible to incorporate temporal structure directly into the regression.  Deciding exactly which quantities to measure and how they should be incorporated, however, are complex questions that are outside the scope of this initial study.

Lastly, capturing the full range of animal behaviors for a large number of animals presents a number of technological challenges, which is why we focused on measuring behavior in a highly simplified environment. However, a more complete understanding of the structure of behavior will require more sophisticated ways to capture behavioral dynamics in more naturalistic settings and during complex social arrangements. While modern deep learning methods have made tracking animals in more realistic settings increasingly plausible \cite{Pereira:2019iv,mathis2020}, there are still considerable hurdles to translating this information into a form that can be subjected to the kind of analysis we propose here.

Despite these limitations, this work represents a new way to quantitatively characterize the evolution of complex behaviors, which may provide new phenotypes that can be subjected to experimental analysis. In the absence of a behavioral fossil record, reconstructing ancestral behaviors requires an inferential approach like the one we present here. In addition, more complex models could be built to test assumptions underlying this initial, diffusion-based, model. Finally, a strength of our approach is that it makes falsifiable predictions about how behaviors are linked mechanistically, providing predictions that can be tested experimentally to provide further insight in the genetic and neurobiological structure of behavior.

\section*{Materials and Methods}

\subsection*{Data collection} All imaging of fly behavior followed the procedures described in \cite{cande2018optogenetic}, but without any red light stimulation.  In total, we collected data from 561 individual from 18 strains and six species.  These included three strains of \emph{D. mauritiana} (\emph{mau29}: 29 flies, \emph{mau317}: 35 flies, \emph{mau318}: 32 flies), four strains of \emph{D. melanogaster} (\emph{Canton-S}: 31 flies, \emph{Oregon-R}: 33 flies, \emph{mel54}: 34 flies, \emph{mel56}: 31 flies), three strains of \emph{D. santomea} (\emph{san00}: 29 flies, \emph{san1482}: 33 flies, \emph{STO OBAT}: 22 flies), three strains of \emph{D. sechellia} (\emph{sech28}: 32 flies, \emph{sech340}: 25 flies, \emph{sech349}: 33 flies), three strains of \emph{D. simulans} (\emph{sim5}: 33 flies, \emph{sim199}: 30 flies, \emph{Oxnard}: 34 flies), and two strains of \emph{D. yakuba} (\emph{yak01}: 34 flies, \emph{CYO2}: 31 flies).

\subsection*{Generalized Linear Mixed Model} We fit our GLMM  (Eq.~\ref{Model1}) using the software introduced in \cite{Hadfield2010}.  The covariance matrices $ V^{(e)}$ and $V^{(a)}\in\Re^{K\times K}$, $K=134$ and the mean vector  $\vec{\mu}\in\Re^{K\times 1}$, were inferred from the posterior distribution via MCMC sampling.  Prior distributions for the covariance matrices were given by Inverse Wishart Distributions (conjugate priors for the multi-Gaussian model) with $K$ degrees of freedom and $\frac{1}{K+1}\frac{I+J}{2}$ as scale matrix, with $J$ and $I$ the unit and identity matrices respectively.  Tree branch length were estimated from \cite{4z}.

\subsection*{Gelman-Rubin convergence diagnostic} This test evaluates MCMC convergence by analyzing the difference between several Markov chains. Convergence is evaluated by comparing the estimated between-chains and within-chain variances for each parameter of the model. Large differences between these variances indicate non-convergence  \cite{gelmanrubin92}. Let $\theta$ be the model parameter of interest and $\left\lbrace\theta_{m}\right\rbrace_{t=1}^{N}$ be the $m$th simulated chain, $m=1,2,...,M$. Denote, $\hat{\theta}_{m}$ and $\hat{\sigma}_{m}^{2}$ be the sample posterior mean and variance of the $m$th chain. If $\hat\theta= \frac{1}{M}\sum_{m=1}^{M}\hat{\theta}_{m}$ is the overall posterior mean estimator, the between-chains and within-chain variances are given by:
\begin{equation}
B=\frac{N}{M-1}\sum_{m=1}^{M}(\hat\theta_{m}-\hat\theta)^{2}\mbox{, } W=\frac{1}{M}\sum_{m=1}^{M}\hat\sigma_{m}^{2}.
\label{BetCh}
\end{equation}
In reference \cite{gelmanrubin92}, it is shown that the following weighted average of $W$ and $B$ is an unbiased estimator of the marginal posterior variance of $\theta$:
$\hat{V}=\frac{N-1}{N}W+\frac{M+1}{NM}B$.

The ratio $\hat{V}/W$ should get close to one as the M chains converge to the target distribution with $N \to \infty$. In reference \cite{BrooksGel97} this ratio known as the potential scale reduction factor (PSRF) was corrected to account for the the sampling variability using $R_{c}=\sqrt{\frac{d+3}{d+1}\frac{\hat{V}}{W}}$,
where $d$ is the degrees of freedom estimate of a t-distribution. Values of PSRF for all model parameters such that $R_{c}<1.1$ are used in \cite{BrooksGel97} as a criteria for convergence of the MCMC chains. Here, we used 20 independent chains, each with a different initialization.

\subsection*{Information-based clustering}
Minimizes the distance between elements within clusters while compressing the original representation. The method minimizes the functional $\mathcal{F}=\langle d \rangle + T I(C;i)$, where $I(C;i)=\sum_{i=1}^{N}\sum_{C=1}^{N_{c}}P(C ; i) \log [\frac{P(C \mid i)}{P(C)}]$ is the mutual information between the original behavioral variable $i$ and the representation $C$. $\langle d \rangle= \sum_{C=1}^{N_{c}}P(C)d(C)$, and $d(C)$ is the average distance of elements chosen out of a single cluster:
\begin{equation}
d(C)=\sum_{i_{1}}^{N}\sum_{i_{2}}^{N}P(i_{1}\mid C) P(i_{2}\mid C)d(i_{1},i_{2}),
\label{AvgDist}
\end{equation}
with $d(i_{1},i_{2})$ being the distance measure between a pair of elements and $P(i\mid C)$ being the probability to find element $i$ in cluster $C$.

Given $\mid C \mid=N_{c}$, $T$ and a random initial condition for $P(C\mid i)$, a solution is obtained by iterating the following self-consistent equations until the criteria $\frac{\mathcal{F}_{t}-\mathcal{F}_{t+1}}{\mathcal{F}_{t}}<10^{-5}$ is satisfied.  We chose $40,000$ different random values of $T \in [0.1,1000]$,  $N_{c}$ between $2$ and $20$, and performed the optimization in each case until the convergence criterion was met. We defined the Pareto front as the set of solutions $P(C \mid i)$ such that no other solution presents a smaller $\langle d \rangle$ and a smaller $I(C;i)$. Finally, for each number of clusters we selected the solution with the lowest $\langle d \rangle$. 

For each number of clusters, the significance of the optimal value found for $\langle d \rangle$ is shown by comparing it to the average distance corresponding to random cluster assignments. These assignments are made in such a way that the amount of elements per cluster is conserved by randomly shuffling the vector that assigns each behavior to a particular cluster. The values presented in the main text correspond to the mean and standard deviation of $\langle d \rangle$ over $50$ different random trials.

\subsection*{Deterministic Information Bottleneck}
Here we use the Deterministic Information Bottleneck (DIB) method to find coarse-grainings of the behavioral space that optimally predict future states \cite{strouse2017deterministic}.  Inspired by the Information Bottleneck \cite{tishby2000information}, DIB  replaces the compression measure $I(X,Z)$ with the entropy $H(Z)$, thus emphasizing constraints on the representation. DIB minimizes the functional:
\begin{equation}
\mathcal{L_{\alpha}}=H(Z)-\alpha H(Z \mid X) - \beta I(Z;Y),
\label{FunctionalDIB}
\end{equation}
with respect to $p(z\in Z\vert x\in X)$ and takes the limit as $\alpha \to 0$.

To apply DIB to the behavioral dynamics, we count time in units of the transitions between states, providing a discrete time series of behaviors, $b(n)$ that can take on $N=134$ different integer values at each discrete time $n$. Here, we relate the joint distributions of $b(n)$ ($X$ in Eqn.~\ref{FunctionalDIB}) and $b(n+\tau)$ ($Y$) through a coarse-grained clustering of the behavioral states ($Z$). We chose $10,000$ different pairs of random values for $\beta$ between $0.1$ and $10^{4}$ and $N_{c}$ between 2 and 30 clusters. Given $N_{c}$, $\beta$ and a random initial condition for $p(t\mid x)$, we find a solution by iterating the self-consistent equations \cite{strouse2017deterministic} until the convergence criteria $\mid \mathcal{L}^{t}-\mathcal{L}^{t+1}\mid<10^{-6}$ is satisfied. If any cluster has its probability become zero at any iteration, then that cluster is dropped for all future iterations, thus $N_{c}$ is the maximum number of clusters that can be returned. Of these $10,000$ solutions, we keep all solutions that are on the Pareto front (i.e., no other solution has both a higher $I(Y;T)$ and a smaller $H(T)$). The displayed clusters are the solutions on the Pareto front with the largest $I(Y;Z)$ for a given number of clusters.

\subsection*{Weighted Similarity Index}
We quantify the similarity between clustering partitions by calculating the Weighted Similarity Index (WSI), a modification of the Rand Index \cite{rand1971objective} such that behaviors contribute the index according to their overall probability. Specifically,
\begin{equation}
 \begin{aligned}
 \textrm{WSI}&=\sum_{i,j \in S_{a}} \textrm{W}_{ij} + \sum_{k,l \in S_{b}}\textrm{W}_{kl}\mbox{, }
 \textrm{W}_{ij}&=\frac{P_{i}P_{k}}{\sum_{kl}P_{k}P_{l}},
\label{WSI}
 \end{aligned}
\end{equation}
where $S_{a} (S_{b})$ is the set of pairs of behaviors that belong to the same (different) cluster in the two partitions and $P_k$ is the probability of observing behavior $k$.

\begin{acknowledgments}
We thank Ilya Nemenman and Daniel Weissman for their helpful comments on the manuscript. D.G.H. was supported by Programa Raices from the MinCyT.  C.R. was supported by the NSF Physics of Living Systems Student Research Network (1806833). G.J.B. was supported by NIMH R01 MH115831-01, the Human Frontier Science Program (RGY0076/2018), and a Cottrell Scholar Award, a program of the Research Corporation for Science Advancement (25999).  J.C., D.L.S., and G.J.B. were supported by the Howard Hughes Medical Institute and the Janelia visiting researcher program. 
\end{acknowledgments}


\bibliographystyle{apsrev4-1}

\begin{thebibliography}{50}%
\makeatletter
\providecommand \@ifxundefined [1]{%
 \@ifx{#1\undefined}
}%
\providecommand \@ifnum [1]{%
 \ifnum #1\expandafter \@firstoftwo
 \else \expandafter \@secondoftwo
 \fi
}%
\providecommand \@ifx [1]{%
 \ifx #1\expandafter \@firstoftwo
 \else \expandafter \@secondoftwo
 \fi
}%
\providecommand \natexlab [1]{#1}%
\providecommand \enquote  [1]{``#1''}%
\providecommand \bibnamefont  [1]{#1}%
\providecommand \bibfnamefont [1]{#1}%
\providecommand \citenamefont [1]{#1}%
\providecommand \href@noop [0]{\@secondoftwo}%
\providecommand \href [0]{\begingroup \@sanitize@url \@href}%
\providecommand \@href[1]{\@@startlink{#1}\@@href}%
\providecommand \@@href[1]{\endgroup#1\@@endlink}%
\providecommand \@sanitize@url [0]{\catcode `\\12\catcode `\$12\catcode
  `\&12\catcode `\#12\catcode `\^12\catcode `\_12\catcode `\%12\relax}%
\providecommand \@@startlink[1]{}%
\providecommand \@@endlink[0]{}%
\providecommand \url  [0]{\begingroup\@sanitize@url \@url }%
\providecommand \@url [1]{\endgroup\@href {#1}{\urlprefix }}%
\providecommand \urlprefix  [0]{URL }%
\providecommand \Eprint [0]{\href }%
\providecommand \doibase [0]{http://dx.doi.org/}%
\providecommand \selectlanguage [0]{\@gobble}%
\providecommand \bibinfo  [0]{\@secondoftwo}%
\providecommand \bibfield  [0]{\@secondoftwo}%
\providecommand \translation [1]{[#1]}%
\providecommand \BibitemOpen [0]{}%
\providecommand \bibitemStop [0]{}%
\providecommand \bibitemNoStop [0]{.\EOS\space}%
\providecommand \EOS [0]{\spacefactor3000\relax}%
\providecommand \BibitemShut  [1]{\csname bibitem#1\endcsname}%
\let\auto@bib@innerbib\@empty
\bibitem [{\citenamefont {Lorenz}(1958)}]{cid}%
  \BibitemOpen
  \bibfield  {author} {\bibinfo {author} {\bibfnamefont {K.~Z.}\ \bibnamefont
  {Lorenz}},\ }\href {\doibase 10.1038/scientificamerican1258-67} {\bibfield
  {journal} {\bibinfo  {journal} {Scientific American}\ }\textbf {\bibinfo
  {volume} {199}},\ \bibinfo {pages} {67} (\bibinfo {year} {1958})}\BibitemShut
  {NoStop}%
\bibitem [{\citenamefont {Martins}\ and\ \citenamefont
  {Martins}(1996)}]{martins1996phylogenies}%
  \BibitemOpen
  \bibfield  {author} {\bibinfo {author} {\bibfnamefont {E.~P.}\ \bibnamefont
  {Martins}}\ and\ \bibinfo {author} {\bibfnamefont {E.~L.~P.}\ \bibnamefont
  {Martins}},\ }\href@noop {} {\emph {\bibinfo {title} {Phylogenies and the
  comparative method in animal behavior}}}\ (\bibinfo  {publisher} {Oxford
  University Press},\ \bibinfo {year} {1996})\BibitemShut {NoStop}%
\bibitem [{\citenamefont {Baier}\ and\ \citenamefont
  {Hoekstra}(2019)}]{baier2019genetics}%
  \BibitemOpen
  \bibfield  {author} {\bibinfo {author} {\bibfnamefont {F.}~\bibnamefont
  {Baier}}\ and\ \bibinfo {author} {\bibfnamefont {H.~E.}\ \bibnamefont
  {Hoekstra}},\ }\href@noop {} {\bibfield  {journal} {\bibinfo  {journal}
  {Proceedings of the Royal Society B}\ }\textbf {\bibinfo {volume} {286}},\
  \bibinfo {pages} {20191697} (\bibinfo {year} {2019})}\BibitemShut {NoStop}%
\bibitem [{\citenamefont {West-Eberhard}(2003)}]{west2003developmental}%
  \BibitemOpen
  \bibfield  {author} {\bibinfo {author} {\bibfnamefont {M.~J.}\ \bibnamefont
  {West-Eberhard}},\ }\href@noop {} {\emph {\bibinfo {title} {Developmental
  plasticity and evolution}}}\ (\bibinfo  {publisher} {Oxford University
  Press},\ \bibinfo {address} {Oxford, U.K.},\ \bibinfo {year}
  {2003})\BibitemShut {NoStop}%
\bibitem [{\citenamefont {Gleason}\ and\ \citenamefont
  {Ritchie}(2004)}]{gleason2004quantitative}%
  \BibitemOpen
  \bibfield  {author} {\bibinfo {author} {\bibfnamefont {J.~M.}\ \bibnamefont
  {Gleason}}\ and\ \bibinfo {author} {\bibfnamefont {M.~G.}\ \bibnamefont
  {Ritchie}},\ }\href@noop {} {\bibfield  {journal} {\bibinfo  {journal}
  {Genetics}\ }\textbf {\bibinfo {volume} {166}},\ \bibinfo {pages} {1303}
  (\bibinfo {year} {2004})}\BibitemShut {NoStop}%
\bibitem [{\citenamefont {Yamamoto}\ and\ \citenamefont
  {Ishikawa}(2013{\natexlab{a}})}]{yamamoto2013genetic}%
  \BibitemOpen
  \bibfield  {author} {\bibinfo {author} {\bibfnamefont {D.}~\bibnamefont
  {Yamamoto}}\ and\ \bibinfo {author} {\bibfnamefont {Y.}~\bibnamefont
  {Ishikawa}},\ }\href@noop {} {\bibfield  {journal} {\bibinfo  {journal}
  {Journal of Neurogenetics}\ }\textbf {\bibinfo {volume} {27}},\ \bibinfo
  {pages} {130} (\bibinfo {year} {2013}{\natexlab{a}})}\BibitemShut {NoStop}%
\bibitem [{\citenamefont {Ellison}\ \emph {et~al.}(2011)\citenamefont
  {Ellison}, \citenamefont {Wiley},\ and\ \citenamefont
  {Shaw}}]{ellison2011genetics}%
  \BibitemOpen
  \bibfield  {author} {\bibinfo {author} {\bibfnamefont {C.}~\bibnamefont
  {Ellison}}, \bibinfo {author} {\bibfnamefont {C.}~\bibnamefont {Wiley}}, \
  and\ \bibinfo {author} {\bibfnamefont {K.}~\bibnamefont {Shaw}},\ }\href@noop
  {} {\bibfield  {journal} {\bibinfo  {journal} {Journal of Evolutionary
  Biology}\ }\textbf {\bibinfo {volume} {24}},\ \bibinfo {pages} {1110}
  (\bibinfo {year} {2011})}\BibitemShut {NoStop}%
\bibitem [{\citenamefont {Shaw}\ and\ \citenamefont
  {Lesnick}(2009)}]{shaw2009genomic}%
  \BibitemOpen
  \bibfield  {author} {\bibinfo {author} {\bibfnamefont {K.~L.}\ \bibnamefont
  {Shaw}}\ and\ \bibinfo {author} {\bibfnamefont {S.~C.}\ \bibnamefont
  {Lesnick}},\ }\href@noop {} {\bibfield  {journal} {\bibinfo  {journal}
  {Proceedings of the National Academy of Sciences}\ }\textbf {\bibinfo
  {volume} {106}},\ \bibinfo {pages} {9737} (\bibinfo {year}
  {2009})}\BibitemShut {NoStop}%
\bibitem [{\citenamefont {Williams}\ and\ \citenamefont
  {Carroll}(2009)}]{williams2009}%
  \BibitemOpen
  \bibfield  {author} {\bibinfo {author} {\bibfnamefont {T.~M.}\ \bibnamefont
  {Williams}}\ and\ \bibinfo {author} {\bibfnamefont {S.~B.}\ \bibnamefont
  {Carroll}},\ }\href@noop {} {\bibfield  {journal} {\bibinfo  {journal}
  {Nature Reviews Genetics}\ }\textbf {\bibinfo {volume} {10}},\ \bibinfo
  {pages} {797} (\bibinfo {year} {2009})}\BibitemShut {NoStop}%
\bibitem [{\citenamefont {Shubin}\ \emph {et~al.}(2009)\citenamefont {Shubin},
  \citenamefont {Tabin},\ and\ \citenamefont {Carroll}}]{shubin2009deep}%
  \BibitemOpen
  \bibfield  {author} {\bibinfo {author} {\bibfnamefont {N.}~\bibnamefont
  {Shubin}}, \bibinfo {author} {\bibfnamefont {C.}~\bibnamefont {Tabin}}, \
  and\ \bibinfo {author} {\bibfnamefont {S.~B.}\ \bibnamefont {Carroll}},\
  }\href@noop {} {\bibfield  {journal} {\bibinfo  {journal} {Nature}\ }\textbf
  {\bibinfo {volume} {457}},\ \bibinfo {pages} {818} (\bibinfo {year}
  {2009})}\BibitemShut {NoStop}%
\bibitem [{\citenamefont {Levine}\ and\ \citenamefont
  {Davidson}(2005)}]{levine2005gene}%
  \BibitemOpen
  \bibfield  {author} {\bibinfo {author} {\bibfnamefont {M.}~\bibnamefont
  {Levine}}\ and\ \bibinfo {author} {\bibfnamefont {E.~H.}\ \bibnamefont
  {Davidson}},\ }\href@noop {} {\bibfield  {journal} {\bibinfo  {journal}
  {Proceedings of the National Academy of Sciences}\ }\textbf {\bibinfo
  {volume} {102}},\ \bibinfo {pages} {4936} (\bibinfo {year}
  {2005})}\BibitemShut {NoStop}%
\bibitem [{\citenamefont {Stern}\ and\ \citenamefont
  {Frankel}(2013)}]{Stern:2013gv}%
  \BibitemOpen
  \bibfield  {author} {\bibinfo {author} {\bibfnamefont {D.~L.}\ \bibnamefont
  {Stern}}\ and\ \bibinfo {author} {\bibfnamefont {N.}~\bibnamefont
  {Frankel}},\ }\href {\doibase 10.1098/rstb.2013.0028} {\bibfield  {journal}
  {\bibinfo  {journal} {Philosophical Transactions of the Royal Society B:
  Biological Sciences}\ }\textbf {\bibinfo {volume} {368}},\ \bibinfo {pages}
  {20130028} (\bibinfo {year} {2013})}\BibitemShut {NoStop}%
\bibitem [{\citenamefont {Baker}\ \emph {et~al.}(2001)\citenamefont {Baker},
  \citenamefont {Taylor},\ and\ \citenamefont {Hall}}]{Baker:2001gp}%
  \BibitemOpen
  \bibfield  {author} {\bibinfo {author} {\bibfnamefont {B.~S.}\ \bibnamefont
  {Baker}}, \bibinfo {author} {\bibfnamefont {B.~J.}\ \bibnamefont {Taylor}}, \
  and\ \bibinfo {author} {\bibfnamefont {J.}~\bibnamefont {Hall}},\ }\href
  {\doibase 10.1016/s0092-8674(01)00293-8} {\bibfield  {journal} {\bibinfo
  {journal} {Cell}\ }\textbf {\bibinfo {volume} {105}},\ \bibinfo {pages} {13}
  (\bibinfo {year} {2001})}\BibitemShut {NoStop}%
\bibitem [{\citenamefont {Cande}\ \emph {et~al.}(2012)\citenamefont {Cande},
  \citenamefont {Andolfatto}, \citenamefont {Prud'homme}, \citenamefont
  {Stern},\ and\ \citenamefont {Gompel}}]{Jessica2012}%
  \BibitemOpen
  \bibfield  {author} {\bibinfo {author} {\bibfnamefont {J.}~\bibnamefont
  {Cande}}, \bibinfo {author} {\bibfnamefont {P.}~\bibnamefont {Andolfatto}},
  \bibinfo {author} {\bibfnamefont {B.}~\bibnamefont {Prud'homme}}, \bibinfo
  {author} {\bibfnamefont {D.~L.}\ \bibnamefont {Stern}}, \ and\ \bibinfo
  {author} {\bibfnamefont {N.}~\bibnamefont {Gompel}},\ }\href {\doibase
  10.1371/journal.pone.0043888} {\bibfield  {journal} {\bibinfo  {journal}
  {PLoS One}\ }\textbf {\bibinfo {volume} {7}},\ \bibinfo {pages} {1} (\bibinfo
  {year} {2012})}\BibitemShut {NoStop}%
\bibitem [{\citenamefont {Cande}\ \emph {et~al.}(2014)\citenamefont {Cande},
  \citenamefont {Stern}, \citenamefont {Morita}, \citenamefont {Prud’homme},\
  and\ \citenamefont {Gompel}}]{Cande2014}%
  \BibitemOpen
  \bibfield  {author} {\bibinfo {author} {\bibfnamefont {J.}~\bibnamefont
  {Cande}}, \bibinfo {author} {\bibfnamefont {D.~L.}\ \bibnamefont {Stern}},
  \bibinfo {author} {\bibfnamefont {T.}~\bibnamefont {Morita}}, \bibinfo
  {author} {\bibfnamefont {B.}~\bibnamefont {Prud’homme}}, \ and\ \bibinfo
  {author} {\bibfnamefont {N.}~\bibnamefont {Gompel}},\ }\href@noop {}
  {\bibfield  {journal} {\bibinfo  {journal} {Cell reports}\ }\textbf {\bibinfo
  {volume} {8}},\ \bibinfo {pages} {363} (\bibinfo {year} {2014})}\BibitemShut
  {NoStop}%
\bibitem [{\citenamefont {Ding}\ \emph {et~al.}(2019)\citenamefont {Ding},
  \citenamefont {Lillvis}, \citenamefont {Cande}, \citenamefont {Berman},
  \citenamefont {Arthur}, \citenamefont {Xu}, \citenamefont {Dickson},\ and\
  \citenamefont {Stern}}]{YunDing2019}%
  \BibitemOpen
  \bibfield  {author} {\bibinfo {author} {\bibfnamefont {Y.}~\bibnamefont
  {Ding}}, \bibinfo {author} {\bibfnamefont {J.~L.}\ \bibnamefont {Lillvis}},
  \bibinfo {author} {\bibfnamefont {J.}~\bibnamefont {Cande}}, \bibinfo
  {author} {\bibfnamefont {G.~J.}\ \bibnamefont {Berman}}, \bibinfo {author}
  {\bibfnamefont {B.~J.}\ \bibnamefont {Arthur}}, \bibinfo {author}
  {\bibfnamefont {M.}~\bibnamefont {Xu}}, \bibinfo {author} {\bibfnamefont
  {B.~J.}\ \bibnamefont {Dickson}}, \ and\ \bibinfo {author} {\bibfnamefont
  {D.~L.}\ \bibnamefont {Stern}},\ }\href@noop {} {\bibfield  {journal}
  {\bibinfo  {journal} {Current Biology}\ }\textbf {\bibinfo {volume} {29}},\
  \bibinfo {pages} {1089} (\bibinfo {year} {2019})}\BibitemShut {NoStop}%
\bibitem [{\citenamefont {Webert}\ \emph {et~al.}(2013)\citenamefont {Webert},
  \citenamefont {Peterson},\ and\ \citenamefont {Hoekstra}}]{Hoekstra2013}%
  \BibitemOpen
  \bibfield  {author} {\bibinfo {author} {\bibfnamefont {J.~N.}\ \bibnamefont
  {Webert}}, \bibinfo {author} {\bibfnamefont {B.~K.}\ \bibnamefont
  {Peterson}}, \ and\ \bibinfo {author} {\bibfnamefont {H.~E.}\ \bibnamefont
  {Hoekstra}},\ }\href {\doibase 10.1038/nature11816} {\bibfield  {journal}
  {\bibinfo  {journal} {Nature}\ }\textbf {\bibinfo {volume} {493}},\ \bibinfo
  {pages} {402} (\bibinfo {year} {2013})}\BibitemShut {NoStop}%
\bibitem [{\citenamefont {Hu}\ and\ \citenamefont
  {Hoekstra}(2016)}]{Hu:2017:Seminar}%
  \BibitemOpen
  \bibfield  {author} {\bibinfo {author} {\bibfnamefont {C.~K.}\ \bibnamefont
  {Hu}}\ and\ \bibinfo {author} {\bibfnamefont {H.~E.}\ \bibnamefont
  {Hoekstra}},\ }\href {\doibase 10.1016/j.semcdb.2016.08.001} {\bibfield
  {journal} {\bibinfo  {journal} {Seminars in cell \& developmental biology}\
  }\textbf {\bibinfo {volume} {61}},\ \bibinfo {pages} {107} (\bibinfo {year}
  {2016})}\BibitemShut {NoStop}%
\bibitem [{\citenamefont {Berman}\ \emph {et~al.}(2014)\citenamefont {Berman},
  \citenamefont {Choi}, \citenamefont {Bialek},\ and\ \citenamefont
  {Shaevitz}}]{berman2014mapping}%
  \BibitemOpen
  \bibfield  {author} {\bibinfo {author} {\bibfnamefont {G.~J.}\ \bibnamefont
  {Berman}}, \bibinfo {author} {\bibfnamefont {D.~M.}\ \bibnamefont {Choi}},
  \bibinfo {author} {\bibfnamefont {W.}~\bibnamefont {Bialek}}, \ and\ \bibinfo
  {author} {\bibfnamefont {J.~W.}\ \bibnamefont {Shaevitz}},\ }\href@noop {}
  {\bibfield  {journal} {\bibinfo  {journal} {Journal of The Royal Society
  Interface}\ }\textbf {\bibinfo {volume} {11}},\ \bibinfo {pages} {20140672}
  (\bibinfo {year} {2014})}\BibitemShut {NoStop}%
\bibitem [{\citenamefont {Berman}(2018)}]{GordonReview}%
  \BibitemOpen
  \bibfield  {author} {\bibinfo {author} {\bibfnamefont {G.~J.}\ \bibnamefont
  {Berman}},\ }\href {\doibase 10.1186/s12915-018-0494-7} {\bibfield  {journal}
  {\bibinfo  {journal} {BMC Biology}\ }\textbf {\bibinfo {volume} {16}},\
  \bibinfo {pages} {23} (\bibinfo {year} {2018})}\BibitemShut {NoStop}%
\bibitem [{\citenamefont {Brown}\ and\ \citenamefont
  {de~Bivort}(2018)}]{Brown:2018ck}%
  \BibitemOpen
  \bibfield  {author} {\bibinfo {author} {\bibfnamefont {A.~E.~X.}\
  \bibnamefont {Brown}}\ and\ \bibinfo {author} {\bibfnamefont
  {B.}~\bibnamefont {de~Bivort}},\ }\href {\doibase 10.1038/s41567-018-0093-0}
  {\bibfield  {journal} {\bibinfo  {journal} {Nature Physics}\ }\textbf
  {\bibinfo {volume} {20}},\ \bibinfo {pages} {410} (\bibinfo {year}
  {2018})}\BibitemShut {NoStop}%
\bibitem [{\citenamefont {Cande}\ \emph {et~al.}(2018)\citenamefont {Cande},
  \citenamefont {Namiki}, \citenamefont {Qiu}, \citenamefont {Korff},
  \citenamefont {Card}, \citenamefont {Shaevitz}, \citenamefont {Stern},\ and\
  \citenamefont {Berman}}]{cande2018optogenetic}%
  \BibitemOpen
  \bibfield  {author} {\bibinfo {author} {\bibfnamefont {J.}~\bibnamefont
  {Cande}}, \bibinfo {author} {\bibfnamefont {S.}~\bibnamefont {Namiki}},
  \bibinfo {author} {\bibfnamefont {J.}~\bibnamefont {Qiu}}, \bibinfo {author}
  {\bibfnamefont {W.}~\bibnamefont {Korff}}, \bibinfo {author} {\bibfnamefont
  {G.~M.}\ \bibnamefont {Card}}, \bibinfo {author} {\bibfnamefont {J.~W.}\
  \bibnamefont {Shaevitz}}, \bibinfo {author} {\bibfnamefont {D.~L.}\
  \bibnamefont {Stern}}, \ and\ \bibinfo {author} {\bibfnamefont {G.~J.}\
  \bibnamefont {Berman}},\ }\href@noop {} {\bibfield  {journal} {\bibinfo
  {journal} {eLife}\ }\textbf {\bibinfo {volume} {7}},\ \bibinfo {pages}
  {e34275} (\bibinfo {year} {2018})}\BibitemShut {NoStop}%
\bibitem [{\citenamefont {Berman}\ \emph {et~al.}(2016)\citenamefont {Berman},
  \citenamefont {Bialek},\ and\ \citenamefont {Shaevitz}}]{Berman18102016}%
  \BibitemOpen
  \bibfield  {author} {\bibinfo {author} {\bibfnamefont {G.~J.}\ \bibnamefont
  {Berman}}, \bibinfo {author} {\bibfnamefont {W.}~\bibnamefont {Bialek}}, \
  and\ \bibinfo {author} {\bibfnamefont {J.~W.}\ \bibnamefont {Shaevitz}},\
  }\href {\doibase 10.1073/pnas.1607601113} {\bibfield  {journal} {\bibinfo
  {journal} {Proceedings of the National Academy of Sciences}\ }\textbf
  {\bibinfo {volume} {113}},\ \bibinfo {pages} {11943} (\bibinfo {year}
  {2016})}\BibitemShut {NoStop}%
\bibitem [{\citenamefont {Hadfield}(2010)}]{Hadfield2010}%
  \BibitemOpen
  \bibfield  {author} {\bibinfo {author} {\bibfnamefont {J.}~\bibnamefont
  {Hadfield}},\ }\href@noop {} {\bibfield  {journal} {\bibinfo  {journal}
  {Journal of Statistical Software}\ }\textbf {\bibinfo {volume} {33}},\
  \bibinfo {pages} {1} (\bibinfo {year} {2010})}\BibitemShut {NoStop}%
\bibitem [{\citenamefont {Felsenstein}(1985)}]{Felsenstein:1985}%
  \BibitemOpen
  \bibfield  {author} {\bibinfo {author} {\bibfnamefont {J.}~\bibnamefont
  {Felsenstein}},\ }\href {\doibase 10.1086/284325} {\bibfield  {journal}
  {\bibinfo  {journal} {The American Naturalist}\ }\textbf {\bibinfo {volume}
  {125}},\ \bibinfo {pages} {1} (\bibinfo {year} {1985})}\BibitemShut {NoStop}%
\bibitem [{\citenamefont {Hadfield}\ and\ \citenamefont
  {Nakagawa}(2010{\natexlab{a}})}]{y4r}%
  \BibitemOpen
  \bibfield  {author} {\bibinfo {author} {\bibfnamefont {J.~D.}\ \bibnamefont
  {Hadfield}}\ and\ \bibinfo {author} {\bibfnamefont {S.}~\bibnamefont
  {Nakagawa}},\ }\href {\doibase 10.1111/j.1420-9101.2009.01915.x} {\bibfield
  {journal} {\bibinfo  {journal} {Journal of evolutionary biology}\ }\textbf
  {\bibinfo {volume} {23}},\ \bibinfo {pages} {494} (\bibinfo {year}
  {2010}{\natexlab{a}})}\BibitemShut {NoStop}%
\bibitem [{\citenamefont {\emph{Drosophila}
  12~Genomes~Consortium}(2007)}]{Clark:2007gf}%
  \BibitemOpen
  \bibfield  {author} {\bibinfo {author} {\bibnamefont {\emph{Drosophila}
  12~Genomes~Consortium}},\ }\href {\doibase 10.1038/nature06341} {\bibfield
  {journal} {\bibinfo  {journal} {Nature}\ }\textbf {\bibinfo {volume} {450}},\
  \bibinfo {pages} {203 } (\bibinfo {year} {2007})}\BibitemShut {NoStop}%
\bibitem [{\citenamefont {Obbard}\ \emph {et~al.}(2012)\citenamefont {Obbard},
  \citenamefont {Maclennan}, \citenamefont {Kim}, \citenamefont {Rambaut},
  \citenamefont {O{\textquoteright}Grady},\ and\ \citenamefont
  {Jiggins}}]{Obbard:2012bk}%
  \BibitemOpen
  \bibfield  {author} {\bibinfo {author} {\bibfnamefont {D.~J.}\ \bibnamefont
  {Obbard}}, \bibinfo {author} {\bibfnamefont {J.}~\bibnamefont {Maclennan}},
  \bibinfo {author} {\bibfnamefont {K.-W.}\ \bibnamefont {Kim}}, \bibinfo
  {author} {\bibfnamefont {A.}~\bibnamefont {Rambaut}}, \bibinfo {author}
  {\bibfnamefont {P.~M.}\ \bibnamefont {O{\textquoteright}Grady}}, \ and\
  \bibinfo {author} {\bibfnamefont {F.~M.}\ \bibnamefont {Jiggins}},\ }\href
  {\doibase 10.1093/molbev/mss150} {\bibfield  {journal} {\bibinfo  {journal}
  {Molecular Biology and Evolution}\ }\textbf {\bibinfo {volume} {29}},\
  \bibinfo {pages} {3459} (\bibinfo {year} {2012})}\BibitemShut {NoStop}%
\bibitem [{\citenamefont {Chyb}\ and\ \citenamefont
  {Gompel}(2013)}]{chyb2013atlas}%
  \BibitemOpen
  \bibfield  {author} {\bibinfo {author} {\bibfnamefont {S.}~\bibnamefont
  {Chyb}}\ and\ \bibinfo {author} {\bibfnamefont {N.}~\bibnamefont {Gompel}},\
  }\href@noop {} {\emph {\bibinfo {title} {Atlas of Drosophila Morphology:
  Wild-type and classical mutants}}}\ (\bibinfo  {publisher} {Academic Press},\
  \bibinfo {year} {2013})\BibitemShut {NoStop}%
\bibitem [{\citenamefont {Seetharam}\ and\ \citenamefont {Stuart}(2013)}]{4z}%
  \BibitemOpen
  \bibfield  {author} {\bibinfo {author} {\bibfnamefont {A.~S.}\ \bibnamefont
  {Seetharam}}\ and\ \bibinfo {author} {\bibfnamefont {G.~W.}\ \bibnamefont
  {Stuart}},\ }\href {\doibase 10.7717/peerj.226} {\bibfield  {journal}
  {\bibinfo  {journal} {PeerJ}\ }\textbf {\bibinfo {volume} {1}},\ \bibinfo
  {pages} {e226} (\bibinfo {year} {2013})}\BibitemShut {NoStop}%
\bibitem [{\citenamefont {Stern}\ \emph {et~al.}(2017)\citenamefont {Stern},
  \citenamefont {Crocker}, \citenamefont {Ding}, \citenamefont {Frankel},
  \citenamefont {Kappes}, \citenamefont {Kim}, \citenamefont {Kuzmickas},
  \citenamefont {Lemire}, \citenamefont {Mast},\ and\ \citenamefont
  {Picard}}]{Stern:2017:G3}%
  \BibitemOpen
  \bibfield  {author} {\bibinfo {author} {\bibfnamefont {D.~L.}\ \bibnamefont
  {Stern}}, \bibinfo {author} {\bibfnamefont {J.}~\bibnamefont {Crocker}},
  \bibinfo {author} {\bibfnamefont {Y.}~\bibnamefont {Ding}}, \bibinfo {author}
  {\bibfnamefont {N.}~\bibnamefont {Frankel}}, \bibinfo {author} {\bibfnamefont
  {G.}~\bibnamefont {Kappes}}, \bibinfo {author} {\bibfnamefont
  {E.}~\bibnamefont {Kim}}, \bibinfo {author} {\bibfnamefont {R.}~\bibnamefont
  {Kuzmickas}}, \bibinfo {author} {\bibfnamefont {A.}~\bibnamefont {Lemire}},
  \bibinfo {author} {\bibfnamefont {J.~D.}\ \bibnamefont {Mast}}, \ and\
  \bibinfo {author} {\bibfnamefont {S.}~\bibnamefont {Picard}},\ }\href
  {\doibase 10.1534/g3.116.038885} {\bibfield  {journal} {\bibinfo  {journal}
  {G3}\ }\textbf {\bibinfo {volume} {7}},\ \bibinfo {pages} {1339} (\bibinfo
  {year} {2017})}\BibitemShut {NoStop}%
\bibitem [{\citenamefont {Yamamoto}\ and\ \citenamefont
  {Ishikawa}(2013{\natexlab{b}})}]{6cr}%
  \BibitemOpen
  \bibfield  {author} {\bibinfo {author} {\bibfnamefont {D.}~\bibnamefont
  {Yamamoto}}\ and\ \bibinfo {author} {\bibfnamefont {Y.}~\bibnamefont
  {Ishikawa}},\ }\href {\doibase 10.3109/01677063.2013.800060} {\bibfield
  {journal} {\bibinfo  {journal} {Journal of Neurogenetics}\ }\textbf {\bibinfo
  {volume} {27}},\ \bibinfo {pages} {130} (\bibinfo {year}
  {2013}{\natexlab{b}})}\BibitemShut {NoStop}%
\bibitem [{\citenamefont {Auer}\ and\ \citenamefont
  {Benton}(2016)}]{Auer:2016jc}%
  \BibitemOpen
  \bibfield  {author} {\bibinfo {author} {\bibfnamefont {T.~O.}\ \bibnamefont
  {Auer}}\ and\ \bibinfo {author} {\bibfnamefont {R.}~\bibnamefont {Benton}},\
  }\href {\doibase 10.1016/j.conb.2016.01.004} {\bibfield  {journal} {\bibinfo
  {journal} {Current opinion in neurobiology}\ }\textbf {\bibinfo {volume}
  {38}},\ \bibinfo {pages} {18} (\bibinfo {year} {2016})}\BibitemShut {NoStop}%
\bibitem [{\citenamefont {van~der Maaten}\ and\ \citenamefont
  {Hinton}(2008)}]{maaten2008visualizing}%
  \BibitemOpen
  \bibfield  {author} {\bibinfo {author} {\bibfnamefont {L.}~\bibnamefont
  {van~der Maaten}}\ and\ \bibinfo {author} {\bibfnamefont {G.}~\bibnamefont
  {Hinton}},\ }\href@noop {} {\bibfield  {journal} {\bibinfo  {journal}
  {Journal of Machine Learning Research}\ }\textbf {\bibinfo {volume} {9}},\
  \bibinfo {pages} {2579} (\bibinfo {year} {2008})}\BibitemShut {NoStop}%
\bibitem [{\citenamefont {Meyer}(1994)}]{meyer1994topographic}%
  \BibitemOpen
  \bibfield  {author} {\bibinfo {author} {\bibfnamefont {F.}~\bibnamefont
  {Meyer}},\ }\href@noop {} {\bibfield  {journal} {\bibinfo  {journal} {Signal
  processing}\ }\textbf {\bibinfo {volume} {38}},\ \bibinfo {pages} {113}
  (\bibinfo {year} {1994})}\BibitemShut {NoStop}%
\bibitem [{\citenamefont {Goodfellow}\ \emph {et~al.}(2016)\citenamefont
  {Goodfellow}, \citenamefont {Bengio},\ and\ \citenamefont
  {Courville}}]{goodfellow2016deep}%
  \BibitemOpen
  \bibfield  {author} {\bibinfo {author} {\bibfnamefont {I.}~\bibnamefont
  {Goodfellow}}, \bibinfo {author} {\bibfnamefont {Y.}~\bibnamefont {Bengio}},
  \ and\ \bibinfo {author} {\bibfnamefont {A.}~\bibnamefont {Courville}},\
  }\href@noop {} {\emph {\bibinfo {title} {Deep learning}}}\ (\bibinfo
  {publisher} {MIT press},\ \bibinfo {address} {Cambridge, MA},\ \bibinfo
  {year} {2016})\BibitemShut {NoStop}%
\bibitem [{\citenamefont {Yang}\ \emph {et~al.}(2006)\citenamefont {Yang} \emph
  {et~al.}}]{yang2006computational}%
  \BibitemOpen
  \bibfield  {author} {\bibinfo {author} {\bibfnamefont {Z.}~\bibnamefont
  {Yang}} \emph {et~al.},\ }\href@noop {} {\emph {\bibinfo {title}
  {Computational molecular evolution}}},\ Vol.\ \bibinfo {volume} {284}\
  (\bibinfo  {publisher} {Oxford University Press Oxford},\ \bibinfo {address}
  {Oxford, U.K.},\ \bibinfo {year} {2006})\BibitemShut {NoStop}%
\bibitem [{\citenamefont {Cunningham}\ \emph {et~al.}(1998)\citenamefont
  {Cunningham}, \citenamefont {Omland},\ and\ \citenamefont {Oakley}}]{md}%
  \BibitemOpen
  \bibfield  {author} {\bibinfo {author} {\bibfnamefont {C.~W.}\ \bibnamefont
  {Cunningham}}, \bibinfo {author} {\bibfnamefont {K.~E.}\ \bibnamefont
  {Omland}}, \ and\ \bibinfo {author} {\bibfnamefont {T.~H.}\ \bibnamefont
  {Oakley}},\ }\href {\doibase 10.1016/s0169-5347(98)01382-2} {\bibfield
  {journal} {\bibinfo  {journal} {Trends in Ecology \& Evolution}\ }\textbf
  {\bibinfo {volume} {13}},\ \bibinfo {pages} {361} (\bibinfo {year}
  {1998})}\BibitemShut {NoStop}%
\bibitem [{\citenamefont {Hadfield}\ and\ \citenamefont
  {Nakagawa}(2010{\natexlab{b}})}]{hadfield2010general}%
  \BibitemOpen
  \bibfield  {author} {\bibinfo {author} {\bibfnamefont {J.}~\bibnamefont
  {Hadfield}}\ and\ \bibinfo {author} {\bibfnamefont {S.}~\bibnamefont
  {Nakagawa}},\ }\href@noop {} {\bibfield  {journal} {\bibinfo  {journal}
  {Journal of evolutionary biology}\ }\textbf {\bibinfo {volume} {23}},\
  \bibinfo {pages} {494} (\bibinfo {year} {2010}{\natexlab{b}})}\BibitemShut
  {NoStop}%
\bibitem [{\citenamefont {Slonim}\ \emph {et~al.}(2005)\citenamefont {Slonim},
  \citenamefont {Atwal}, \citenamefont {Tkačik},\ and\ \citenamefont
  {Bialek}}]{Slonim20122005}%
  \BibitemOpen
  \bibfield  {author} {\bibinfo {author} {\bibfnamefont {N.}~\bibnamefont
  {Slonim}}, \bibinfo {author} {\bibfnamefont {G.~S.}\ \bibnamefont {Atwal}},
  \bibinfo {author} {\bibfnamefont {G.}~\bibnamefont {Tkačik}}, \ and\
  \bibinfo {author} {\bibfnamefont {W.}~\bibnamefont {Bialek}},\ }\href
  {\doibase 10.1073/pnas.0507432102} {\bibfield  {journal} {\bibinfo  {journal}
  {Proceedings of the National Academy of Sciences}\ }\textbf {\bibinfo
  {volume} {102}},\ \bibinfo {pages} {18297} (\bibinfo {year} {2005})},\
  \Eprint
  {http://arxiv.org/abs/http://www.pnas.org/content/102/51/18297.full.pdf}
  {http://www.pnas.org/content/102/51/18297.full.pdf} \BibitemShut {NoStop}%
\bibitem [{\citenamefont {Tinbergen}(1951)}]{tinbergen51}%
  \BibitemOpen
  \bibfield  {author} {\bibinfo {author} {\bibfnamefont {N.}~\bibnamefont
  {Tinbergen}},\ }\href@noop {} {\emph {\bibinfo {title} {{The Study of
  Instinct}}}}\ (\bibinfo  {publisher} {Oxford University Press},\ \bibinfo
  {address} {Oxford, U. K.},\ \bibinfo {year} {1951})\BibitemShut {NoStop}%
\bibitem [{\citenamefont {Strouse}\ and\ \citenamefont
  {Schwab}(2017)}]{strouse2017deterministic}%
  \BibitemOpen
  \bibfield  {author} {\bibinfo {author} {\bibfnamefont {D.}~\bibnamefont
  {Strouse}}\ and\ \bibinfo {author} {\bibfnamefont {D.~J.}\ \bibnamefont
  {Schwab}},\ }\href@noop {} {\bibfield  {journal} {\bibinfo  {journal} {Neural
  Computation}\ }\textbf {\bibinfo {volume} {29}},\ \bibinfo {pages} {1611}
  (\bibinfo {year} {2017})}\BibitemShut {NoStop}%
\bibitem [{\citenamefont {Rand}(1971)}]{rand1971objective}%
  \BibitemOpen
  \bibfield  {author} {\bibinfo {author} {\bibfnamefont {W.~M.}\ \bibnamefont
  {Rand}},\ }\href@noop {} {\bibfield  {journal} {\bibinfo  {journal} {Journal
  of the American Statistical association}\ }\textbf {\bibinfo {volume} {66}},\
  \bibinfo {pages} {846} (\bibinfo {year} {1971})}\BibitemShut {NoStop}%
\bibitem [{\citenamefont {Anderson}(2016)}]{anderson2016circuit}%
  \BibitemOpen
  \bibfield  {author} {\bibinfo {author} {\bibfnamefont {D.~J.}\ \bibnamefont
  {Anderson}},\ }\href@noop {} {\bibfield  {journal} {\bibinfo  {journal}
  {Nature Reviews Neuroscience}\ }\textbf {\bibinfo {volume} {17}},\ \bibinfo
  {pages} {692} (\bibinfo {year} {2016})}\BibitemShut {NoStop}%
\bibitem [{\citenamefont {Tajima}(1993)}]{Tajima.1993.Genetics}%
  \BibitemOpen
  \bibfield  {author} {\bibinfo {author} {\bibfnamefont {F.}~\bibnamefont
  {Tajima}},\ }\href@noop {} {\bibfield  {journal} {\bibinfo  {journal}
  {Genetics}\ }\textbf {\bibinfo {volume} {135}},\ \bibinfo {pages} {599}
  (\bibinfo {year} {1993})}\BibitemShut {NoStop}%
\bibitem [{\citenamefont {Pereira}\ \emph {et~al.}(2018)\citenamefont
  {Pereira}, \citenamefont {Aldarondo}, \citenamefont {Willmore}, \citenamefont
  {Kislin}, \citenamefont {Wang}, \citenamefont {Murthy},\ and\ \citenamefont
  {Shaevitz}}]{Pereira:2019iv}%
  \BibitemOpen
  \bibfield  {author} {\bibinfo {author} {\bibfnamefont {T.~D.}\ \bibnamefont
  {Pereira}}, \bibinfo {author} {\bibfnamefont {D.~E.}\ \bibnamefont
  {Aldarondo}}, \bibinfo {author} {\bibfnamefont {L.}~\bibnamefont {Willmore}},
  \bibinfo {author} {\bibfnamefont {M.}~\bibnamefont {Kislin}}, \bibinfo
  {author} {\bibfnamefont {S.~S.~H.}\ \bibnamefont {Wang}}, \bibinfo {author}
  {\bibfnamefont {M.}~\bibnamefont {Murthy}}, \ and\ \bibinfo {author}
  {\bibfnamefont {J.~W.}\ \bibnamefont {Shaevitz}},\ }\href {\doibase
  10.1038/s41592-018-0234-5} {\bibfield  {journal} {\bibinfo  {journal} {Nature
  Methods}\ }\textbf {\bibinfo {volume} {16}},\ \bibinfo {pages} {117}
  (\bibinfo {year} {2018})}\BibitemShut {NoStop}%
\bibitem [{\citenamefont {Mathis}\ and\ \citenamefont
  {Mathis}(2020)}]{mathis2020}%
  \BibitemOpen
  \bibfield  {author} {\bibinfo {author} {\bibfnamefont {M.~W.}\ \bibnamefont
  {Mathis}}\ and\ \bibinfo {author} {\bibfnamefont {A.}~\bibnamefont
  {Mathis}},\ }\href {\doibase 10.1016/j.conb.2019.10.008} {\bibfield
  {journal} {\bibinfo  {journal} {Current Opinion in Neurobiology}\ }\textbf
  {\bibinfo {volume} {60}},\ \bibinfo {pages} {1} (\bibinfo {year}
  {2020})}\BibitemShut {NoStop}%
\bibitem [{\citenamefont {Gelman}\ and\ \citenamefont
  {Rubin}(1992)}]{gelmanrubin92}%
  \BibitemOpen
  \bibfield  {author} {\bibinfo {author} {\bibfnamefont {A.}~\bibnamefont
  {Gelman}}\ and\ \bibinfo {author} {\bibfnamefont {D.~B.}\ \bibnamefont
  {Rubin}},\ }\href@noop {} {\bibfield  {journal} {\bibinfo  {journal}
  {Statistical Science}\ ,\ \bibinfo {pages} {457}} (\bibinfo {year}
  {1992})}\BibitemShut {NoStop}%
\bibitem [{\citenamefont {Brooks}\ and\ \citenamefont
  {Gelman}(1998)}]{BrooksGel97}%
  \BibitemOpen
  \bibfield  {author} {\bibinfo {author} {\bibfnamefont {S.~P.}\ \bibnamefont
  {Brooks}}\ and\ \bibinfo {author} {\bibfnamefont {A.}~\bibnamefont
  {Gelman}},\ }\href@noop {} {\bibfield  {journal} {\bibinfo  {journal}
  {Journal of computational and graphical statistics}\ }\textbf {\bibinfo
  {volume} {7}},\ \bibinfo {pages} {434} (\bibinfo {year} {1998})}\BibitemShut
  {NoStop}%
\bibitem [{\citenamefont {Tishby}\ \emph {et~al.}(1999)\citenamefont {Tishby},
  \citenamefont {Pereira},\ and\ \citenamefont
  {Bialek}}]{tishby2000information}%
  \BibitemOpen
  \bibfield  {author} {\bibinfo {author} {\bibfnamefont {N.}~\bibnamefont
  {Tishby}}, \bibinfo {author} {\bibfnamefont {F.~C.}\ \bibnamefont {Pereira}},
  \ and\ \bibinfo {author} {\bibfnamefont {W.}~\bibnamefont {Bialek}},\ }in\
  \href@noop {} {\emph {\bibinfo {booktitle} {Proceedings of the 37th Annual
  Allerton Conference on Communication, Control and Computing}}}\ (\bibinfo
  {publisher} {University of Illinois Press},\ \bibinfo {address}
  {Urbana-Champaign, IL},\ \bibinfo {year} {1999})\ pp.\ \bibinfo {pages}
  {368--377}\BibitemShut {NoStop}%
\end{thebibliography}
%



\renewcommand{\thetable}{S\arabic{table}}
\setcounter{figure}{0}
\renewcommand{\thefigure}{S\arabic{figure}}
\setcounter{table}{0}

\begin{figure*}
\centering
\includegraphics[width=1.\linewidth]{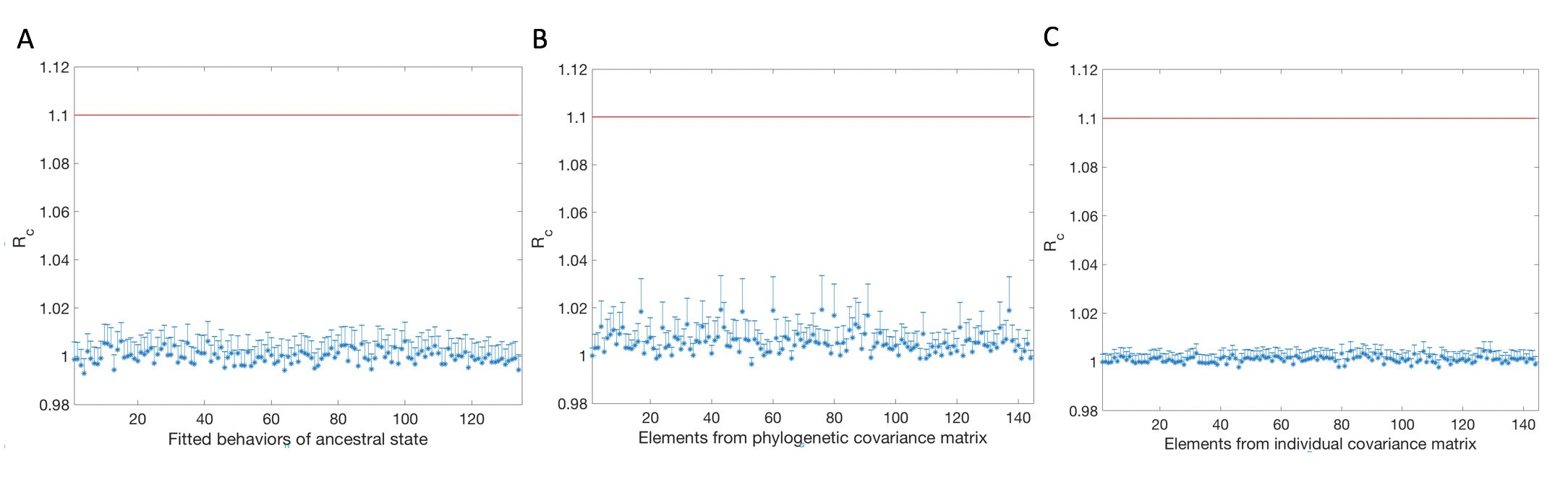}
\caption{Gelman Rubin diagnostic for model parameters inferred using MCMC. A: PSRF for the $134$ ancestral behaviors inferred in the GLMM. $20$ MCMC chains with different initial conditions were used. B: PSRF for the phylogenetic covariance matrix elements corresponding to the $10\%$ most common behaviors performed by the measured flies. C: PSRF for the individual covariance matrix elements corresponding to the $10\%$ most common behaviors performed by the measured flies. The PSRF values for all of these inferred parameters indicate that the MCMC chains are converging.}
\label{RubinGelman}
\end{figure*}

\begin{figure*}
\centering
\includegraphics[width=1.\linewidth]{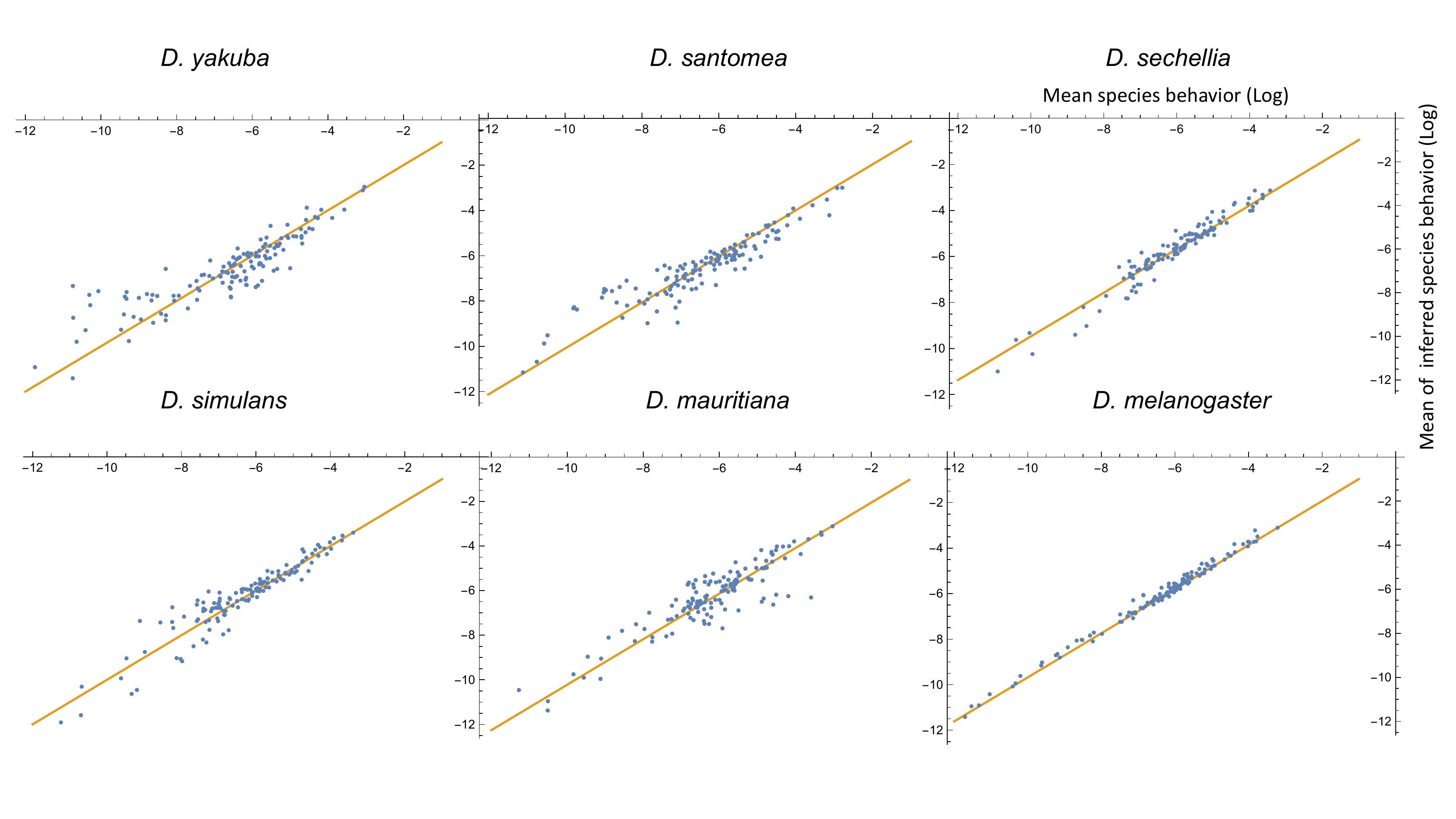}
\caption{Comparison between measured and inferred behaviors (in log scale) for each of the extant species. The mean of the measured behavioral repertoires for all the individuals of a particular species is taken in the log scale. Each measured behavioral mean gets compared to the mean obtained from the components of the MCMC samples corresponding to that particular species and behavioral mode (i.e., the inferred behavioral repertories from the GLMM). The biggest differences occur mostly in the low probability behaviors.}
\label{RubinGelman2}
\end{figure*}

\begin{figure*}
\centering
\includegraphics[width=1.\linewidth]{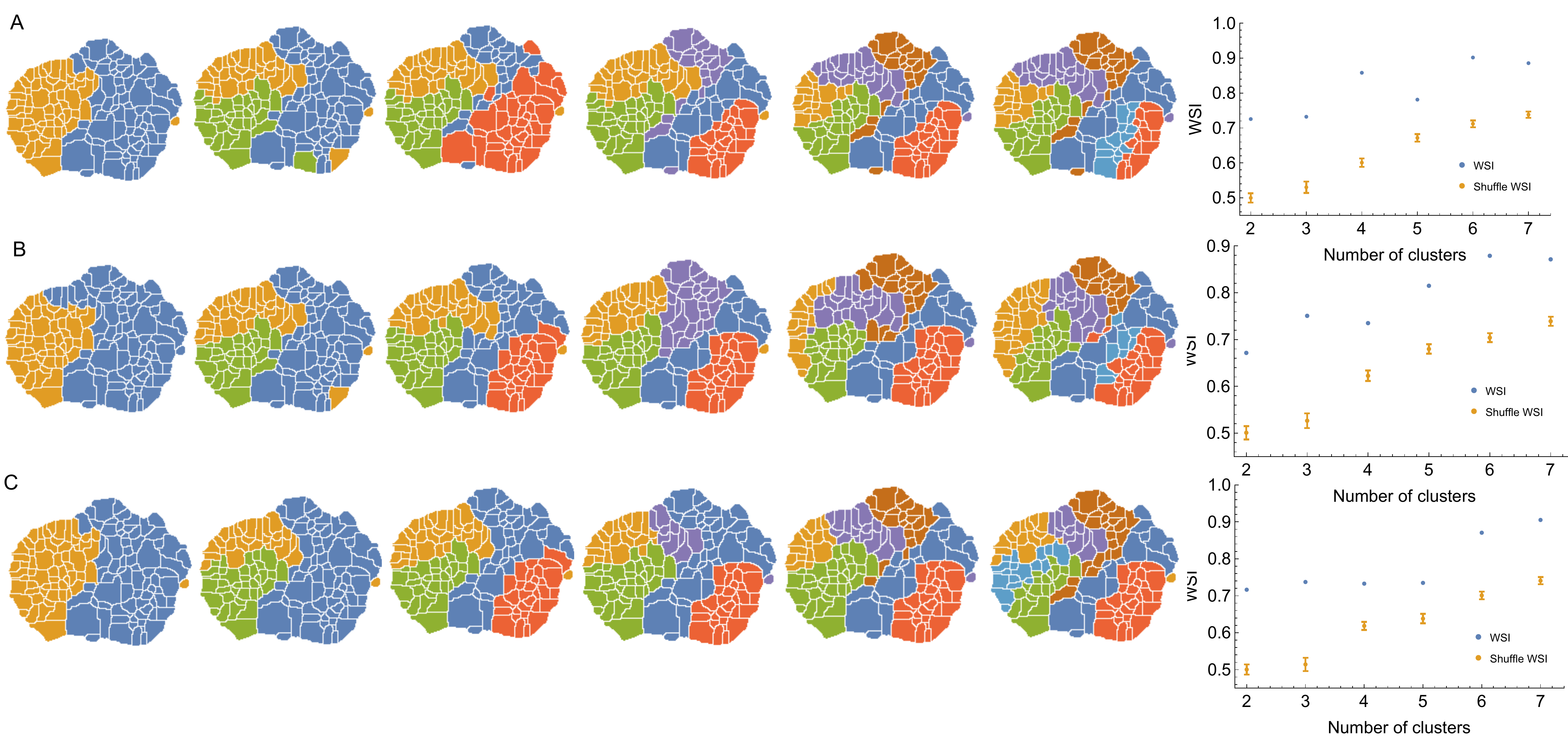}
\caption{Behaviors clustered according to information of the individual covariance matrix using three different clustering methods. A: Results using k-medoids clustering method with  distance matrix $d_{ij}=(1-\rho_{ij})/2$ for 2,3,..7 clusters. To the right, the WSI between the clusters obtained using k-medoids and those obtained using predictive information bottleneck method. Clearly, the similarity between these two orthogonal measurements is significant, as can be shown when compared to the WSI calculated by randomly shuffling the labels of the k-medoids clustering corresponding to each number of clusters. B: Same as in A but we used Spectral clustering instead of k-medoids. The similarity index between Spectral clustering and predictive information bottleneck is also statistically significant. C: Same as in A but we used Information based clustering instead of k-medoids. The similarity index between Information based clustering and predictive information bottleneck is statistically significant as well.}
\label{ClusterTypes}
\end{figure*}

\begin{figure*}
\centering
\includegraphics[width=1.\linewidth]{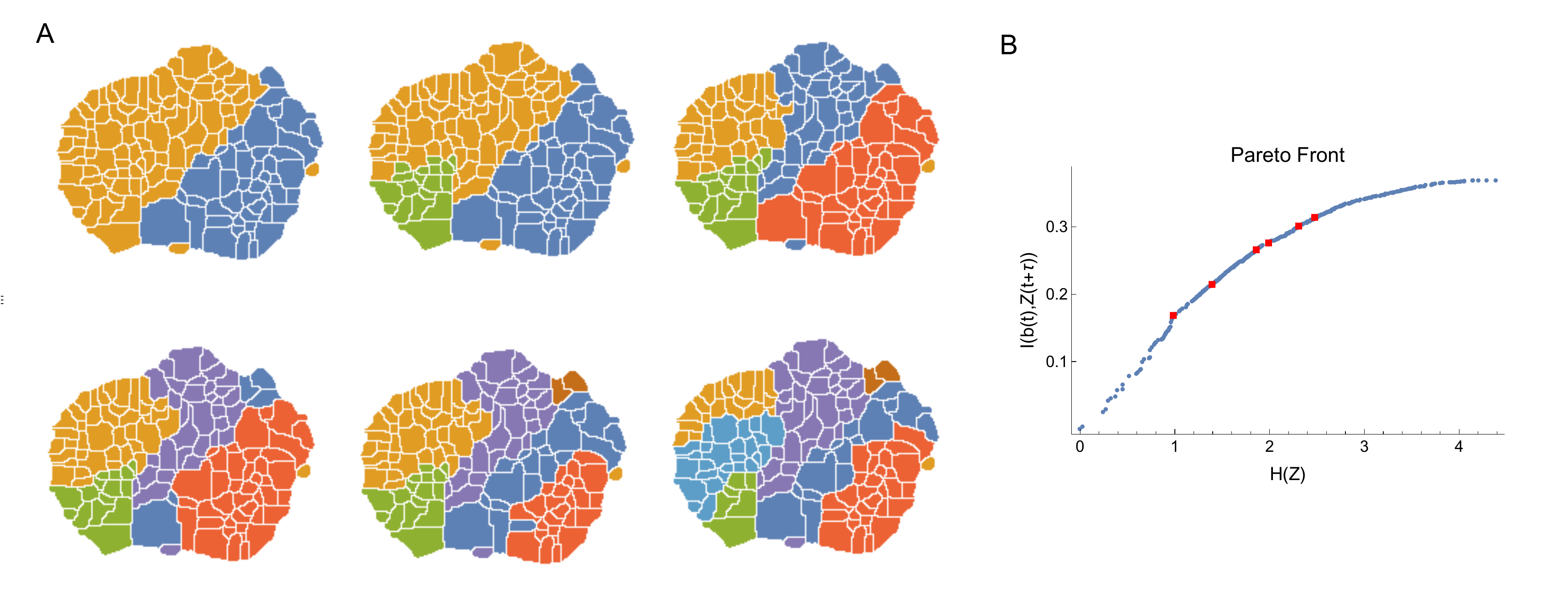}
\caption{Coarse-grained behavioral representations that are optimally predictive of the future behavior states via DIB. A: Behavioral representation with 2,3,...,7 clusters using $\tau=50$ in Eq.~\ref{FunctionalDIB}. B: Optimal trade-off curve (Pareto Front) between complexity of coarse grained description against predictive power. For each number of clusters, representations in A correspond to points (in red) in this curve with the highest predictive information}
\label{ParetoFrontDIB}
\end{figure*}

\begin{figure*}
\centering
\includegraphics[width=0.8\linewidth]{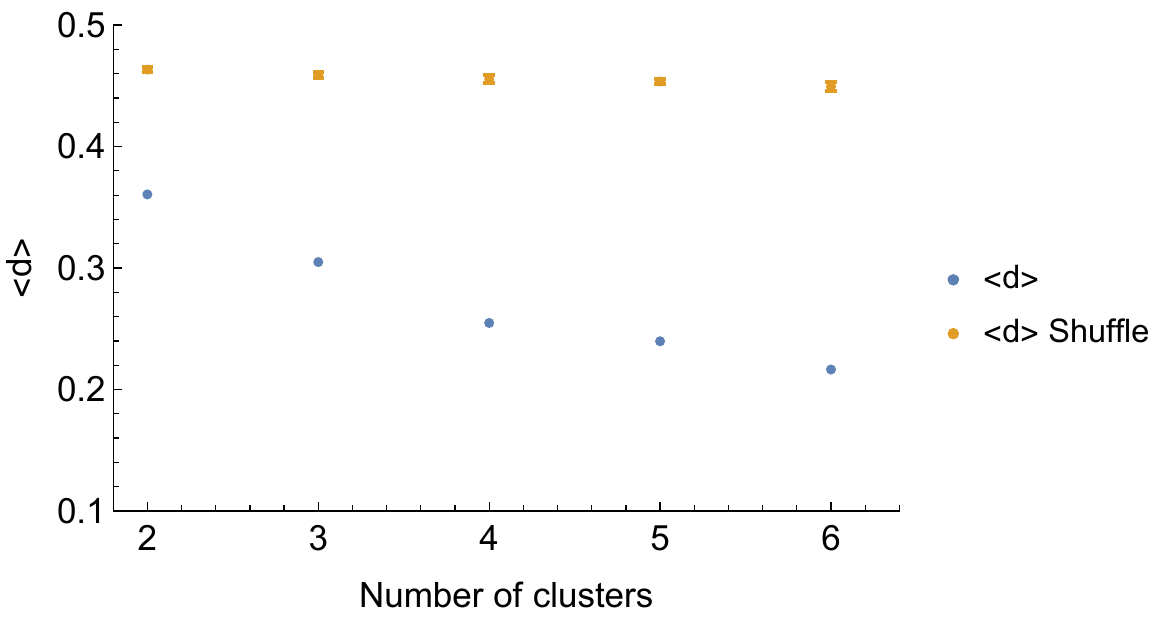}
\caption{Modularity measure of the intra-species behavioral covariance matrix using information based clustering. $<d>$ corresponds to the average distance among elements of the same clusters, (see Materials and Methods for definition). We show for different number of clusters that matrix modularity is significantly smaller (in blue) than expected by random assignation of behaviors to clusters (in orange).   }
\label{ShuffleInfo}
\end{figure*}

\end{document}